\documentclass{appolb}
\def\sun{\ast{\odot}}
\def\aap{A\&A\,  }
\def\aj{AJ  }
\def\apj{ApJ\,  }
 
\def\mnras{MNRAS\,  }

\def\za{Z. Astrophys.  } 

\usepackage {graphicx}

\begin{document}
\title{
The Luminosity Function of Galaxies as 
modelled by the Generalized Gamma Distribution
}
\author{Lorenzo Zaninetti
\address{Dipartimento  di Fisica Generale,
 via P.Giuria 1,\\ I-10125 Turin,Italy}
}
\maketitle
\begin{abstract}
Two new luminosity functions of galaxies can be
built starting from three and four parameter 
generalized gamma distributions.
In the astrophysical conversion, the number of parameters 
increases by one, due to the addition 
of the overall density of galaxies.
A third new galaxy luminosity function is built
starting from 
a three parameter generalized gamma distribution
for the mass of galaxies once 
a simple nonlinear relationship
between mass and luminosity is assumed;
in this case the number of parameters is five
because the overall density of galaxies 
and a parameter that regulates mass and luminosity are 
added.
The three new galaxy luminosity functions were tested  
on the Sloan Digital Sky Survey (SDSS) in five different
bands; the results always produce a "better fit"
than the Schechter function.
The formalism that has been developed allows to analyze the
Schechter function with a transformation of location. 
A test between theoretical and observed number 
of galaxies as a function of redshift 
was done on data extracted from
a two-degree field galaxy redshift survey.
\end{abstract}

\section{Introduction}
The luminosity function of galaxies
$ \Phi (L) dL$ (LF)  
is the number of galaxies per unit
volume, $Mpc^3$, whose luminosity is comprised between 
$L$ and $L+dl$, see Section 3.7 in \cite{Longair2008}
or Section 1.12 in \cite{Padmanabhan_III_2002}.
The luminosity, $L$, has physical units of 
 $Watt~Hz^{-1}$ and  therefore the band being considered
should always be specified.
In our case, we selected the data 
of the  Sloan Digital Sky Survey (SDSS)    
which has  five bands  
$u^*$ ($\lambda = 3550\AA{})$,
$g^*$ ($\lambda = 4770\AA{})$,
$r^*$ ($\lambda = 6230\AA{})$,
$i^*$ ($\lambda = 7620\AA{})$
and 
$z^*$ ($\lambda = 9130\AA{})$
with  $\lambda$ denoting the wavelength of the CCD camera,
see \cite{Gunn1998}.

The Schechter LF of galaxies, 
see \cite{schechter},
is the more widely used function 
\begin{equation}
\Phi (L) dL = (\frac {\Phi^*}{L^*}) (\frac {L}{L^*})^{\alpha}
\exp \bigl ( {- \frac {L}{L^*}} \bigr ) dL \quad,
\label{equation_schechter}
\end {equation}
where $\alpha$ sets the slope for low values 
of $L$, $L^*$ is the
characteristic luminosity and $\Phi^*$ is the normalization.
In the formula above, $ L^* $ characterizes the break
of  the LF. As an example, $L \ge 0.1 L^*$ 
defines a giant  galaxy , see the text after
formula (1.18) in \cite{Gallagher2000}.
An astronomical form of
equation~(\ref{equation_schechter}) can be deduced by 
introducing the
distribution in absolute magnitude
\begin{eqnarray}
\Phi (M)dM=&&(0.4  ln 10) \Phi^* 10^{0.4(\alpha +1 ) (M^*-M)}\nonumber\\
&& \times \exp \bigl ({- 10^{0.4(M^*-M)}} \bigr)  dM \quad,
\label{equation_schechter_M}
\end {eqnarray}
where $M^*$ is the characteristic magnitude as derived from the
data.

This function has three parameters that can be found 
by fitting the data.
Over the years, many modifications have been made
to the Schechter LF in order to improve its fit:
we report two of them.
When the fit of the rich clusters LF
is not satisfactory
a two-component Schechter-like function is introduced,
see~\cite{Driver1996}
\begin{eqnarray}
L_{max} > L > L_{Dwarf} : \quad
\Phi (L) dL = (\frac {\Phi^*}{L^*}) (\frac {L}{L^*})^{\alpha}
\exp \bigl ( {- \frac {L}{L^*}} \bigr ) dL \quad,
\nonumber \\
 \\
L_{Dwarf} > L > L_{min} : \quad
\Phi (L) dL = (\frac {\Phi_{Dwarf}}{L^*}) (\frac {L}{L_{Dwarf}})^{\alpha_{
Dwarf}}
 dL \quad,
\nonumber
\end {eqnarray}
where
\begin{eqnarray}
\Phi_{Dwarf} = \Phi^* (\frac {L_{Dwarf} }{L^*})^{\alpha}
\exp \bigl ( {- \frac {L_{Dwarf} }{L^*}} \bigr )
\quad.
\nonumber
\end{eqnarray}
This two-component function defined between
the maximum luminosity, $L_{max}$,
 and the minimum luminosity, $L_{min}$,
has five parameters because two additional parameters
have been added:
$L_{Dwarf}$ which represents the magnitude where
dwarfs first dominate over giants and ${\alpha_{Dwarf}}$
which regulates 
the faint slope parameter for the dwarf population.

Another LF introduced in order to
fit the case
of extremely low luminosity galaxies
is the double Schechter function with five
parameters, see~\cite{Blanton_2005}:
\begin{equation}
\Phi(L) dL = \frac{dL}{L_\ast}
 \exp(-L/L_{\ast}) \left[
\phi_{\ast,1}
\left( \frac{L}{L_{\ast}} \right)^{\alpha_1} +
\phi_{\ast,2}
\left( \frac{L}{L_{\ast}} \right)^{\alpha_2}
 \right]
\quad,
\end{equation}
where the parameters $\Phi^*$ and $\alpha$ which characterize
the Schechter function have been doubled in $\phi_{\ast,1}$
and $\phi_{\ast,2}$.
The strong dependence of LF 
on different environments such as voids, 
superclusters and supercluster cores was 
analyzed by \cite{Einasto_2009} with
\begin{equation}
F (L) \mathrm{d}L \propto (L/L^{*})^\alpha (1 +
(L/L^{*})^\gamma)^{(\delta-\alpha)/\gamma}
\mathrm{d}(L/L^{*}),
\label{eq:abell}
\end{equation}
where $\alpha$ is the exponent 
at low luminosities $(L/L^{*}) \ll 1$,
$\delta$ is the exponent at high luminosities 
$(L/L^{*}) \gg 1$,
$\gamma$ is a parameter of transition between the
two power laws, and $L^{*}$ is the characteristic luminosity.
The previous LFs 
leave a series of questions
unanswered or partially answered:
\begin {itemize}
\item What is the function of introducing a 
 transformation of location in the LF? 
\item Is it possible to model the data with just 
 a single LF?
\item Is it possible to deduce a LF for galaxies starting
 from the mass distribution of galaxies?
\item Is it possible to improve the 
 Schechter function by introducing a 
 transformation of location? 
\item Does the new LF match the observed 
 behavior of the number of galaxies 
 for a given solid angle and flux
 as a function of the redshift? 
\end {itemize}
In Section~\ref{gammagene},
this paper explores
how a generalized gamma distribution can model two galaxy LFs.
The method which allows the LF to be deduced for galaxies starting
from a mass distribution as given by a generalized gamma 
is presented in Section~\ref{masses}.
Section~\ref{location} reports the analytical and
numerical results on the Schechter function with 
transformation of location.
In Section~\ref{secz}, the redshift dependence
of the Schechter function and one 
of the four new LFs are compared
with data from the
two-degree Field Galaxy Redshift Survey 
in the 2dFGRS catalogue.

\section{The generalized gamma distribution }

\label{gammagene}

The generalized gamma distribution can be represented 
with three parameters, see \cite{Tanemura2003,Hinde1980} 
or four parameters, see \cite{evans}.
We will explore both cases in the following.
In order to make a comparison between our LF 
and the Schechter 
LF, we first down-loaded
the data of the LF of galaxies
in the five bands of SDSS 
adopted in \cite{Blanton_2003}; 
they are available at:
http://cosmo.nyu.edu/blanton/lf.html. 
In the previous paper, \cite{Blanton_2003},
the basic assumption was to consider a Friedmann-Robertson-Walker
cosmological world model with matter density 
$\Omega_0 = 0.3$, vacuum
pressure $\Omega_\Lambda = 0.7$ 
and Hubble constant $H_0 = 100$
$h$ km s$^{-1}$ Mpc$^{-1}$ with $h=1$.
The data contain the absolute magnitude, the value
of the LF for that magnitude and the error
of the LF.

The LF of galaxies as obtained from astronomical observations
ranges in magnitude from a minimum value, $M_{min}$,
to a maximum value, $M_{max}$; details can be found in
\cite{lin} and \cite{Machalski2000}.
A nonlinear fit
through the Levenberg--Marquardt method (subroutine
MRQMIN in \cite{press}) allows the determination
of the parameters, but the first derivative of the
LF with respect to the unknown parameters
should be provided.
The merit function $\chi^2$
can be computed as
\begin{equation}
\chi^2 =
\sum_{j=1}^n ( \frac {LF_{theo} - LF_{astr} } {\sigma_{LF_{astr}}})^2
\quad ,
\label{chisquare}
\end{equation}
where $n$ is number of data and the two 
 indices $theo$ and $astr$ stand for theoretical 
and astronomical, respectively. 

Particular attention should be paid to the number of  
unknown parameters
in the LF:  
three for the  Schechter function 
(formula~(\ref{equation_schechter_M})),
four  for  formula~ (\ref{pdf4magni}), 
four  for the Schechter function 
with transformation of location (formula~(\ref{schechter_loc4})),
five as represented by formula~(\ref{pdfml5})
and 
five for formula~(\ref{pdf5magni}).
A reduced  merit function $\chi_{red}^2$
can be computed as
\begin{equation}
\chi_{red}^2 = \chi^2/NF
\quad,
\label{chisquarereduced}
\end{equation}
where $NF=n-k$, $n$ is the number of data 
and $k$ is the number of parameters.
The Akaike information criterion ($AIC$), see \cite{Akaike1974} ,
is defined as
\begin{equation}
AIC  = 2k - 2  ln(L)
\quad ,
\end {equation}
where $L$ is
the likelihood  function  and $k$  the number of  free parameters
in the model.
We assume  a Gaussian distribution for  the errors
and  the likelihood  function
can be derived  from the $\chi^2$ statistic
$L \propto \exp (- \frac{\chi^2}{2} ) $
where  $\chi^2$ has been computed by  equation~(\ref{chisquare}),
see~\cite{Liddle2004}, \cite{Godlowski2005}.
Now $AIC$ becomes
\begin{equation}
AIC  = 2k + \chi^2
\quad.
\label{AIC}
\end {equation}

The Bayesian information criterion ($BIC$), see \cite{Schwarz1978},
is
\begin{equation}
BIC  = k~ ln(n) - 2  ln(L)
\quad,
\end {equation}
where $L$ is
the likelihood  function, $k$  the number of  free parameters
in the model and  $n$ the number of observations.
The phrase  "better fit" used in the following means that 
the three statistical indicators : $\chi^2$,
$AIC$ and $BIC$ are smaller for the considered LF 
than for the Schechter function .

\subsection{The generalized gamma distribution with five parameters}

The starting point is the 
probability density function (in the following PDF)
named generalized gamma 
that we report exactly as in 
\cite{evans}:
\begin{equation}
G(x;a,b,c,k)=
\frac 
{
{\it k}\, \left( {\frac {x-a}{b}} \right) ^{c{\it k}-1}{{\rm e}^{-
 \left( {\frac {x-a}{b}} \right) ^{{\it k}}}}
}
{
b\Gamma \left( c \right) 
}
\label{base5}
\quad ,
\end{equation}
where 
$\Gamma$ is the gamma function, 
$a$ is the location parameter, 
$b$ is the scale parameter,
$c$ and $k$ are two shape parameters.
The number of parameters is four and
the astrophysical version of the previous PDF can
be obtained by inserting $a=L_a$ , $x=L$ and $b=L^*$:
\begin{equation}
\Psi(L;L^*,c,k,\Psi^*)=
\Psi^*\frac 
{
{\it k}\, \left( {\frac {L-L_a}{L^*}} \right) ^{c{\it k}-1}{{\rm e}^{-
 \left( {\frac {L-L_a}{L^*}} \right) ^{{\it k}}}}
}
{
L^*\Gamma \left( c \right) 
}
\quad ,
\label{lf5}
\end{equation}
where $\Psi^*$ is a normalization factor which defines the
overall density of galaxies, expressed as a number per cubic $Mpc$.
The mathematical range of existence is $ L_a \leq L < \infty $
and the number of parameters is five because 
$\Psi^*$ has been added.
The averaged luminosity, $ { \langle L \rangle } $, is: 
\begin{equation}
{ \langle L \rangle } 
=
\frac
{
{\it L^*}\,\Gamma \left( {\frac {1+ck}{k}} \right) +{\it L_a}\,
\Gamma \left( c \right) 
}
{
\Gamma \left( c \right) 
}
\quad ,
\label{lmedio5}
\end{equation}
and the mode is
at 
\begin{equation}
L=
\left( {\frac {ck-1}{k}} \right) ^{\frac{1}{k}} {\it L^*}+{\it 
L_a}
\quad. 
\label{mode5}
\end{equation}
The relationships connecting the absolute magnitude $M$,
$M^*$ and $M_a$ 
 of a
galaxy to its luminosity are: 
\begin{equation}
\frac {L}{L_{\sun}} =
10^{0.4(M_{\sun} - M)}
\quad,
~\frac {L^*}{L_{\sun}} =
10^{0.4(M_{\sun} - M^*)}
\quad,
~\frac {L_a}{L_{\sun}} =
10^{0.4(M_{\sun} - M_a)}
\quad,
\label{mbolo}
\end {equation}
where $M_{\sun}$ is the absolute magnitude 
of the sun in the considered band.
As an example, the SDSS bands have 
$M_{\sun}$= 4.48 in $z^*$ and 
$M_{\sun}$= 6.32 in $u^*$,
see \cite{blanton}.
A more convenient form of the LF in terms of the absolute
magnitude $M$ is:
\begin{eqnarray}
\Psi (M) dM =
{\it \Psi^*} \frac
{
 0.4\,k\,\ln \left( 10 \right) \left( {10}^{ 0.4\,{\it M^*}- 0.4\,M}-{10}^{ 0.4\,{
\it M^*}- 0.4\,{\it M_a}} \right) ^{ck-1}
}
{
\Gamma \left( c \right) 
}\times \nonumber \\
{{\rm e}^{- \left( {10}
^{ 0.4\,{\it M^*}- 0.4\,M}-{10}^{ 0.4\,{\it M^*}- 0.4\,{\it 
M_a}} \right) ^{k}}} {10}^{ 0.4\,{\it M^*}- 0.4\,M
} 
\,dM \quad.
\label{pdf5magni}
\end {eqnarray}
The mode when expressed in magnitude is at 
\begin{equation}
M = - 1.085\,\ln \left( \left( {\frac {ck- 1}{k}} \right) ^{\frac{1}{k}}
{{\rm e}^{- 0.921\,{\it M^*}}}+{{\rm e}^{-
 0.921\,{\it M_a}}} \right) 
\quad .
\label{mode5magni}
\end{equation}

This data-oriented function contains the five parameters $c$,
$k$, 
$M^*$, 
$M_a$ 
and $\Psi^*$ which can be derived from the 
operation of fitting
observational data.
The results are reported in Table~\ref{dataLF5} 
together with the number of elements $N$ belonging to the sample,
$M_{min}$ and $M_{max}$ of the sample,
the merit function $\chi^2$ and $\chi_{red}^2$,
the $\chi^2$ and $\chi_{red}^2$ of the 
Schechter function as computed by us 
and
the $\chi^2$ of the 
Schechter function as computed by \cite{Blanton_2003}.

 \begin{table}
 \caption[]{
 Parameters of fits to LF
 in SDSS Galaxies of 
 five parameter function represented by 
 formula (\ref{pdf5magni}).
}
 \label{dataLF5} 
 \[
 \begin{array}{lccccc}
 \hline
parameter    &  u^*   &  g^* & r^*  & i^* & z^*  \\ \noalign{\smallskip}
 \hline
 \noalign{\smallskip}
 M_a  ~  & -14.25  
 & -15.55 
 & -15.77  
 & -16.52 
 & -17.39 
 \\ 
M^* - 5\log_{10}h ~   
 & -17.22  
 & -18.29 
 & -20     
 & -20.01 
 & -20.25  
 \\
\Psi^* [h^3~Mpc^{-3}] 
 & 0.078 
 & 0.077  
 & 0.11  
 & 0.05  
 & 0.055 
 \\
c & 0.61 
 & 0.30   
 & 0.19  
 & 0.47  
 & 0.49  
 \\ 
k & 0.82  
 & 0.92 
 & 0.87  
 & 0.78  
 & 0.76  
 \\ 
N & 483
 & 599
 & 674 
 & 709 
 & 740 
 \\
M_{min}  & -20.65
 & -22.09
 & -22.94
 & -23.42
 & -23.73
 \\
M_{max}  & -15.78
 & -16.32
 & -16.30
 & -17.21
 & -17.48
 \\
\chi^2 
 & 282
 & 736
 & 1766
 & 1726 
 & 2136
 \\
\chi_{red}^2 
 & 0.59 
 & 1.24
 & 2.64
 & 2.45
 & 2.90
 \\
AIC~k=5  
 & 292 
 & 746 
 & 1776
 & 1736 
 & 2146
\\
BIC~k=5  
 & 313 
 & 768 
 & 1799
 & 1759 
 & 2169
 \\
\chi^2 -Schechter 
 & 330 
 & 753 
 & 2260
 & 2282 
 & 3245
\\
\chi_{red}^2 -Schechter 
 & 0.689
 & 1.263
 & 3.368
 & 3.232
 & 4.403
\\

\chi^2 -Blanton~2003 
 & 341 
 & 756 
 & 2276
 & 2283 
 & 3262
\\
AIC~k=3 -Schechter
 & 336
 & 759 
 & 2266
 & 2288 
 & 3253
\\
BIC~k=3 -Schechter 
 & 349
 & 772 
 & 2279
 & 2302 
 & 3265
\\
 \end{array}
 \]
 \end {table}

\begin{figure}
 \centering
\includegraphics[width=10cm]{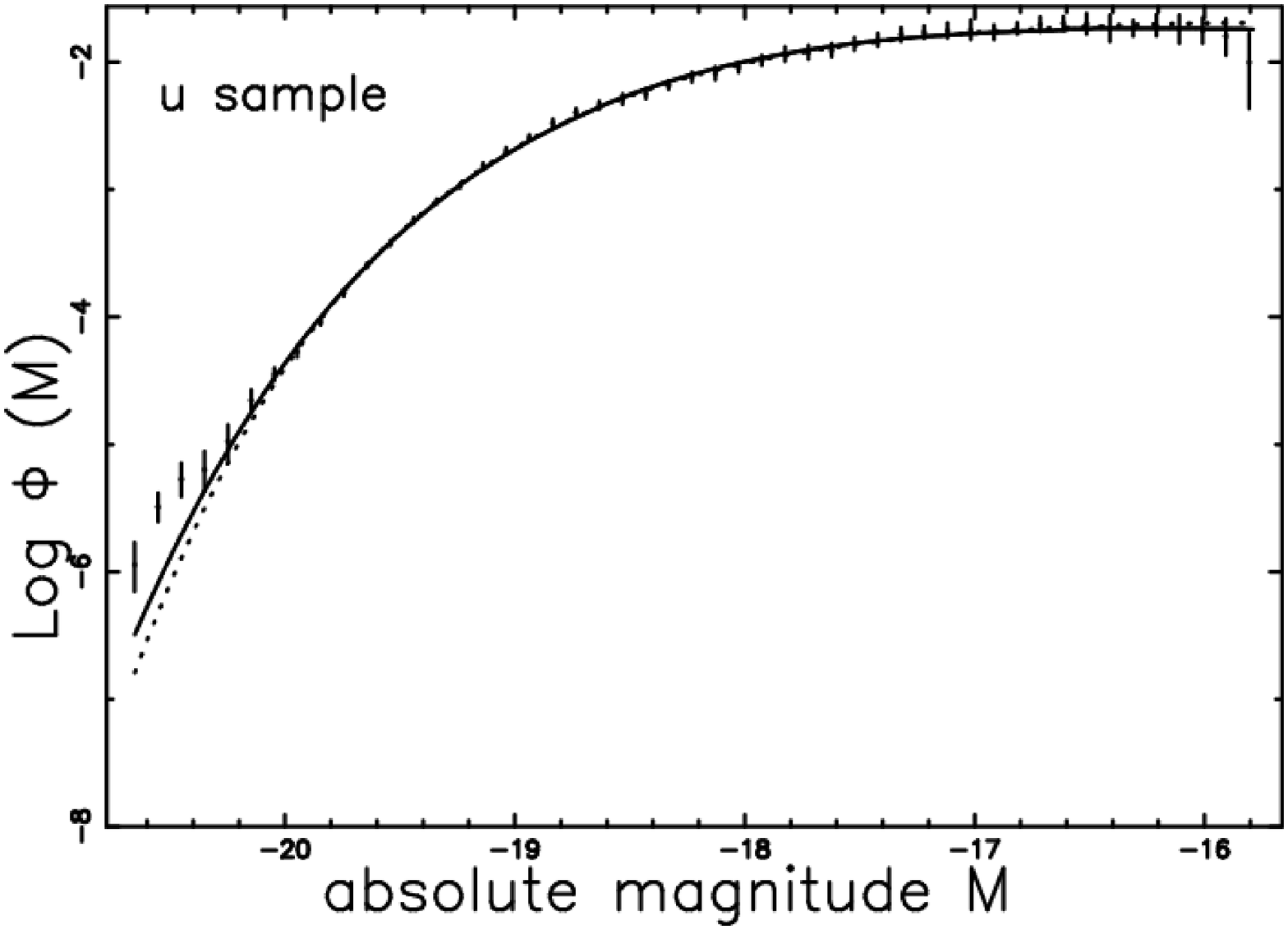}
\caption {The luminosity function data of 
SDSS($u^*$) are represented with error bars. 
The continuous line fit represents our LF
(\ref{pdf5magni}) 
and the dotted 
line represents the Schechter function.
 }
 \label{due_u} 
 \end{figure}

\begin{figure}
 \centering
\includegraphics[width=10cm]{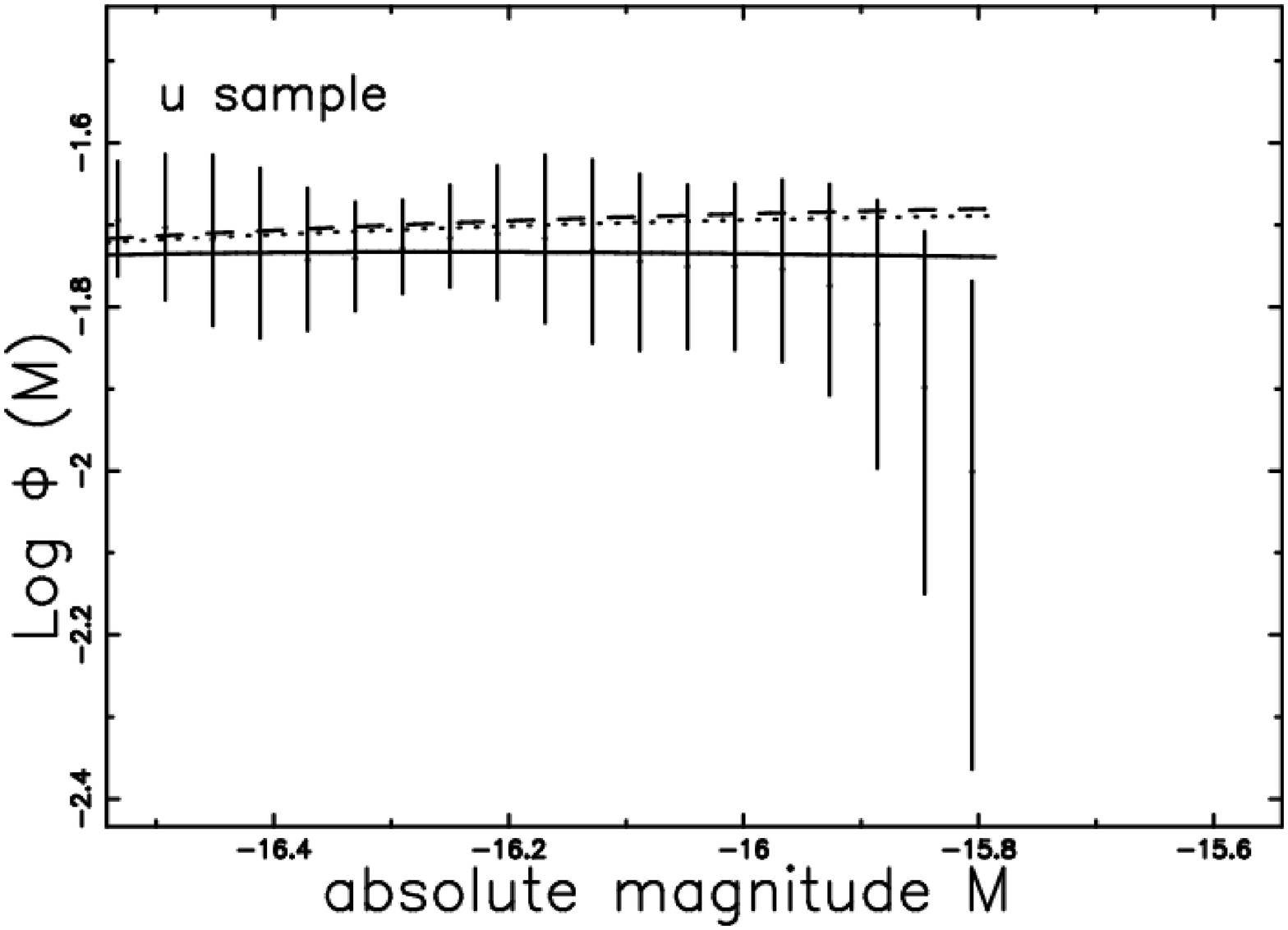}
\caption {The luminosity function data of 
SDSS($u^*$) 
in the low luminosity region 
are represented by the error bar. 
The continuous line fit represents our LF
(\ref{pdf5magni}),
the dotted line represents the Schechter function as 
given by our data
and
the dashed
line represents the Schechter function as given 
by \cite{Blanton_2003}.
 }
 \label{due_low_u} 
 \end{figure}

\begin{figure}
 \centering
\includegraphics[width=10cm]{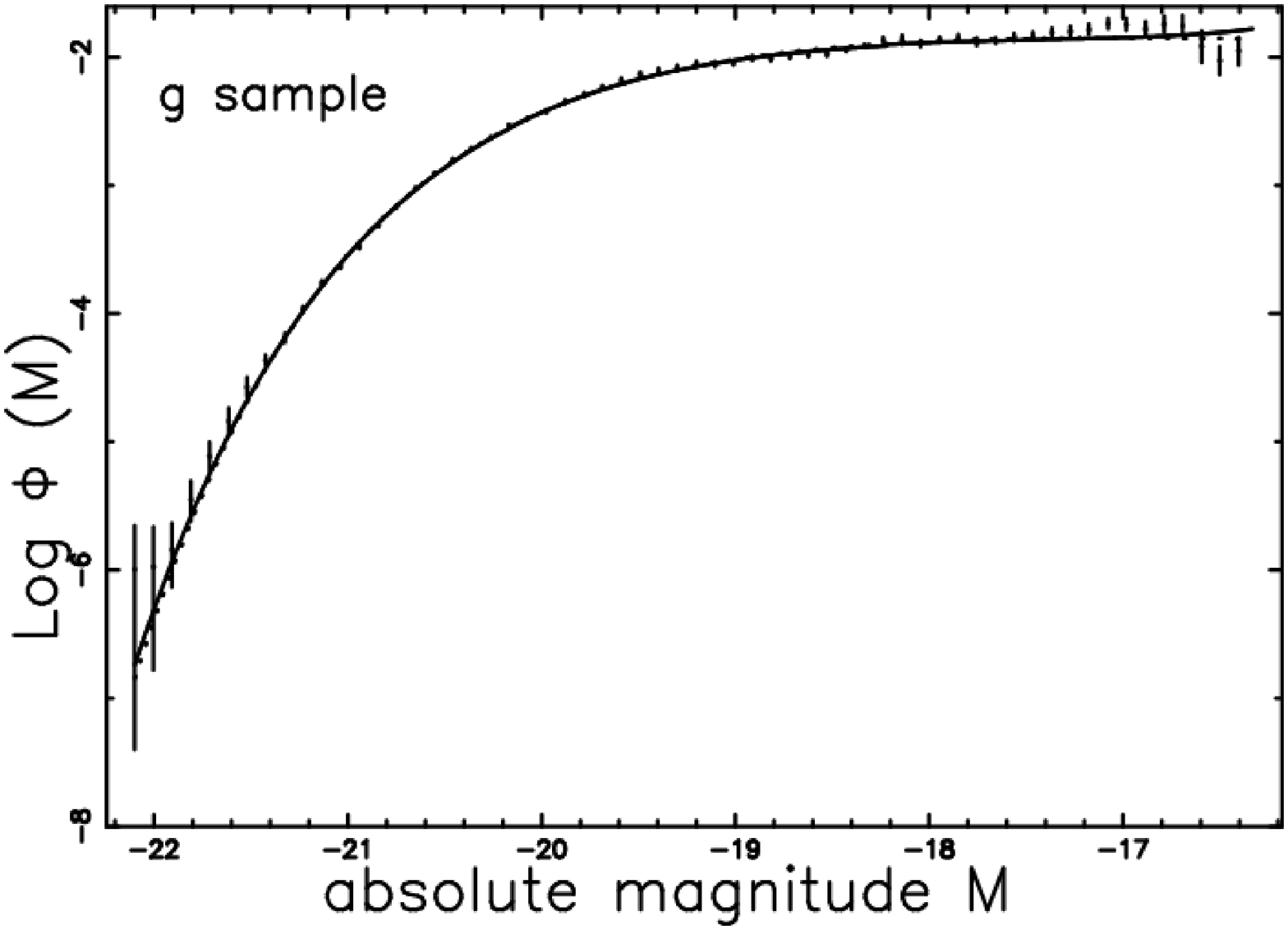}
\caption {The luminosity function data of 
SDSS($g^*$) are represented with error bars. 
The continuous line fit represents our LF~(\ref{pdf5magni}) 
and the dotted 
line represents the Schechter function.
 }
 \label{due_g} 
 \end{figure}

\begin{figure}
 \centering
\includegraphics[width=10cm]{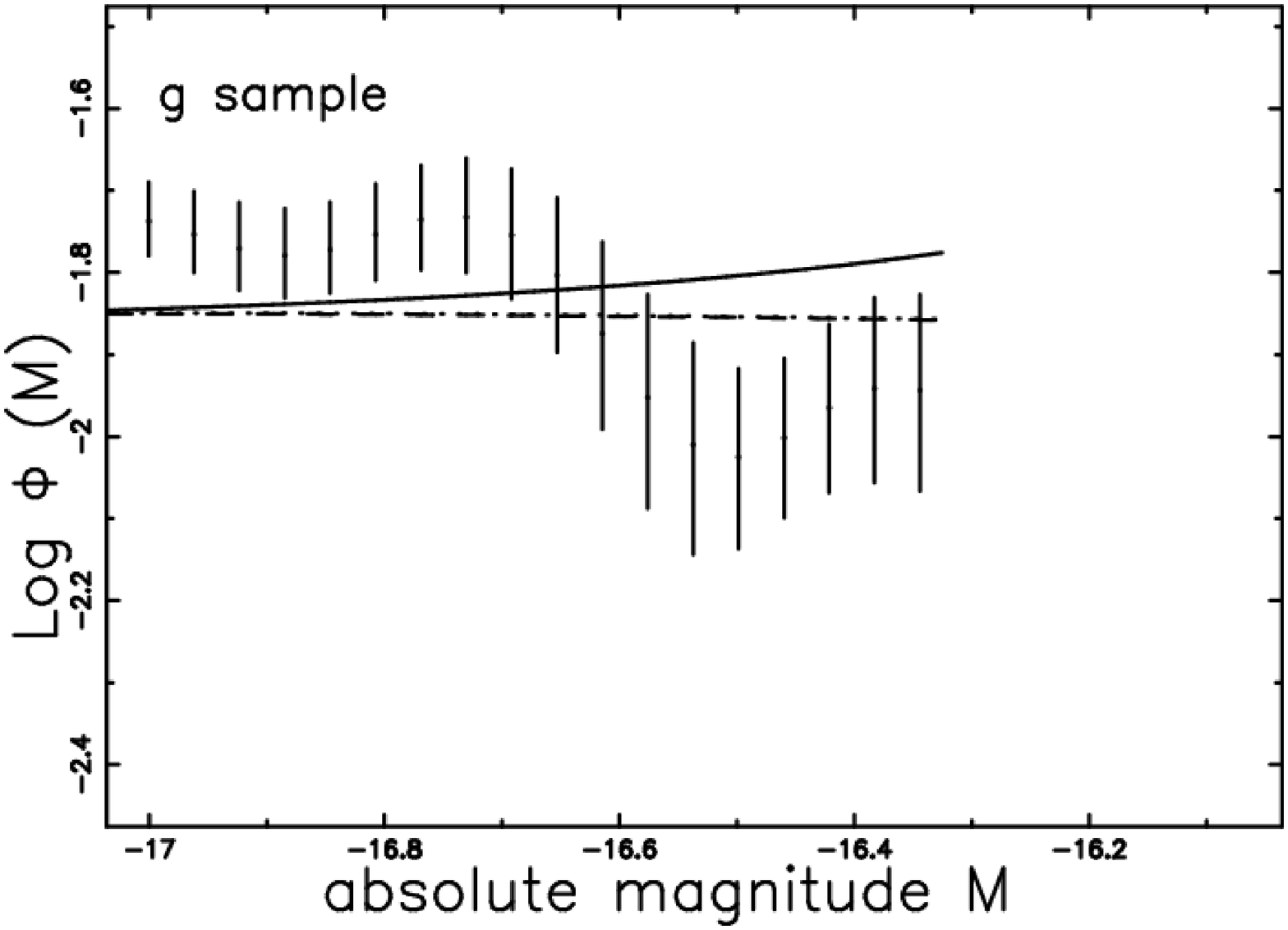}
\caption {The luminosity function data of 
SDSS($g^*$) 
in the low luminosity region 
are represented by the error bar. 
The continuous line fit represents our LF~(\ref{pdf5magni}), 
the dotted line represents the Schechter function as 
given by our data
and
the dashed
line represents the Schechter function as given 
by \cite{Blanton_2003}.
 }
 \label{due_low_g} 
 \end{figure}

\begin{figure}
 \centering
\includegraphics[width=10cm]{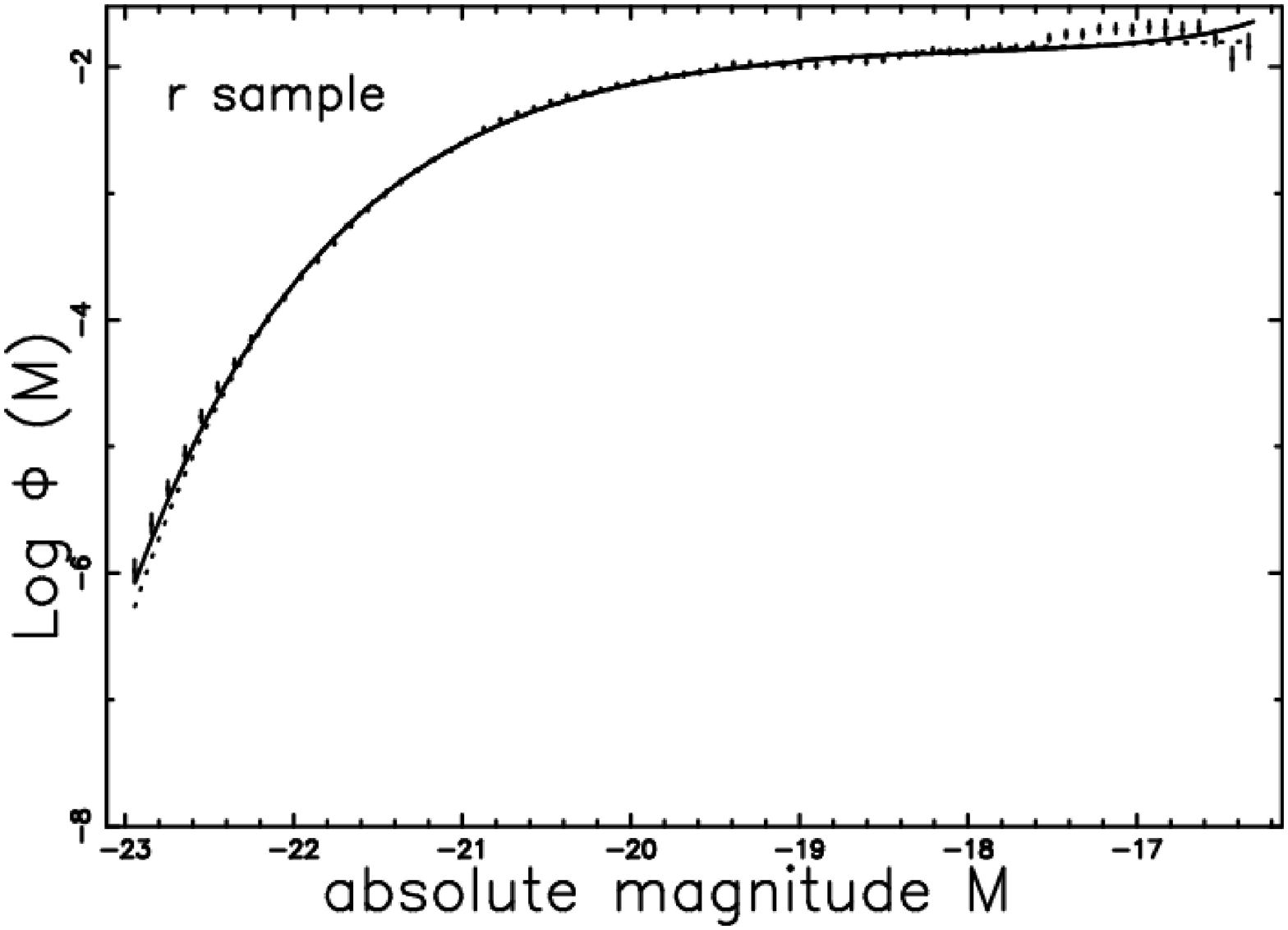}
\caption {The luminosity function data of 
SDSS($r^*$) are represented with error bars. 
The continuous line fit represents our LF~(\ref{pdf5magni}) 
and the dotted 
line represents the Schechter function.
 }
 \label{due_r} 
 \end{figure}

\begin{figure}
 \centering
\includegraphics[width=10cm]{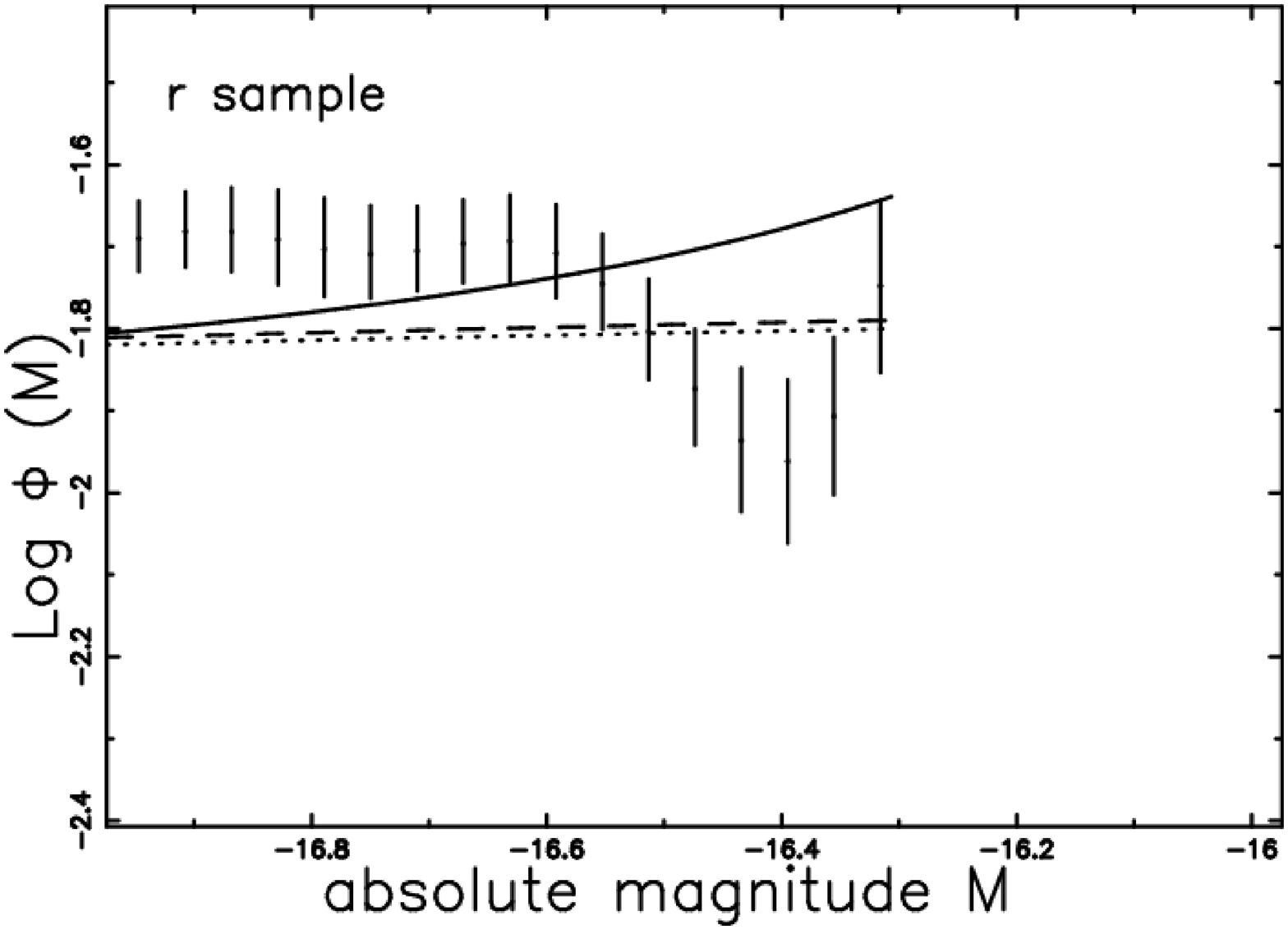}
\caption {The luminosity function data of 
SDSS($r^*$) 
in the low luminosity region 
are represented with error bars. 
The continuous line fit represents our LF~(\ref{pdf5magni}),
the dotted line represents the Schechter function as 
given by our data
and
the dashed
line represents the Schechter function as given 
by \cite{Blanton_2003}.
 }
 \label{due_low_r} 
 \end{figure}

\begin{figure}
 \centering
\includegraphics[width=10cm]{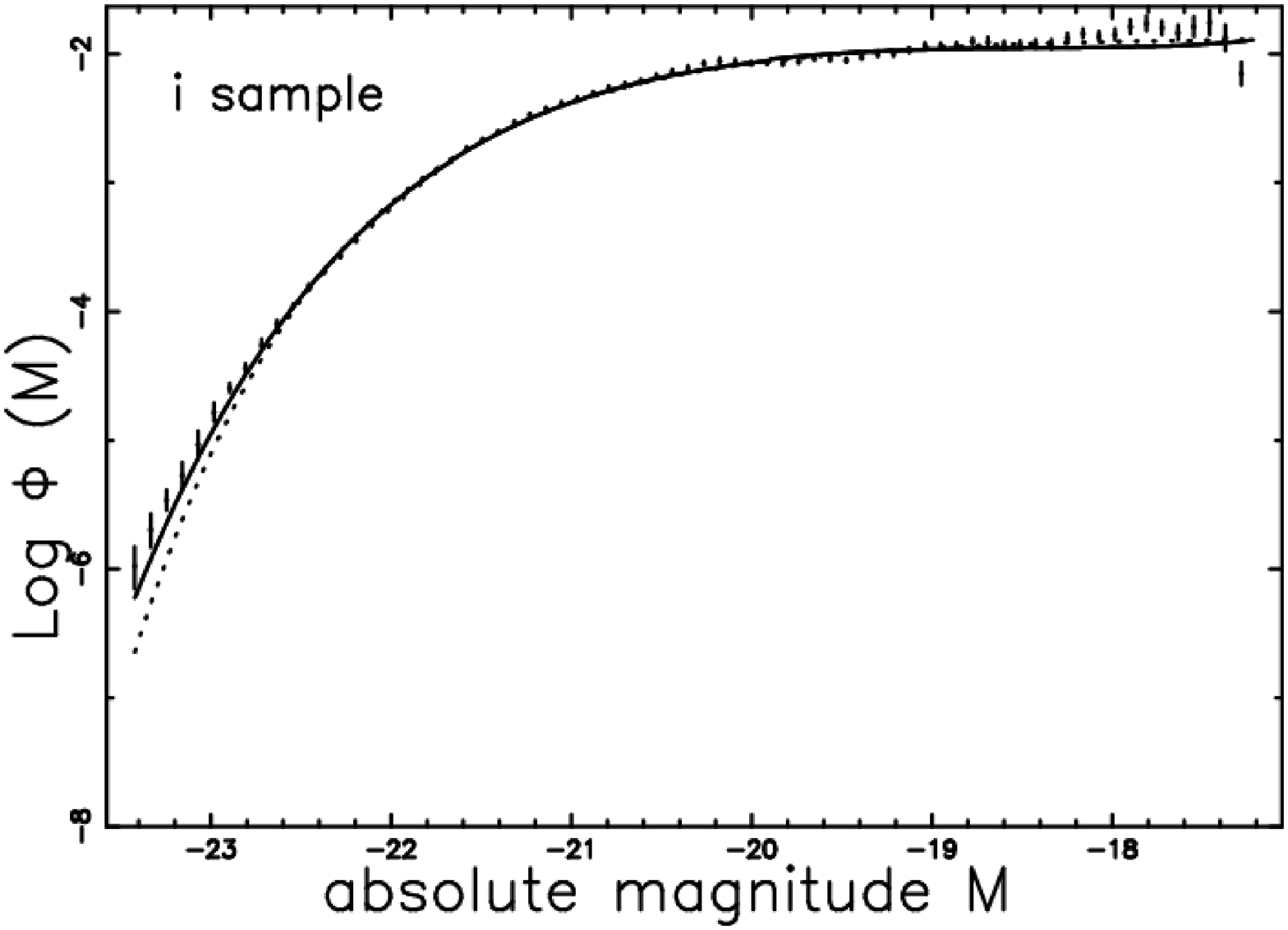}
\caption {The luminosity function data of 
SDSS($i^*$) are represented with error bars. 
The continuous line fit represents our LF~(\ref{pdf5magni}) 
and the dotted 
line represents the Schechter function.
 }
 \label{due_i} 
 \end{figure}

\begin{figure}
 \centering
\includegraphics[width=10cm]{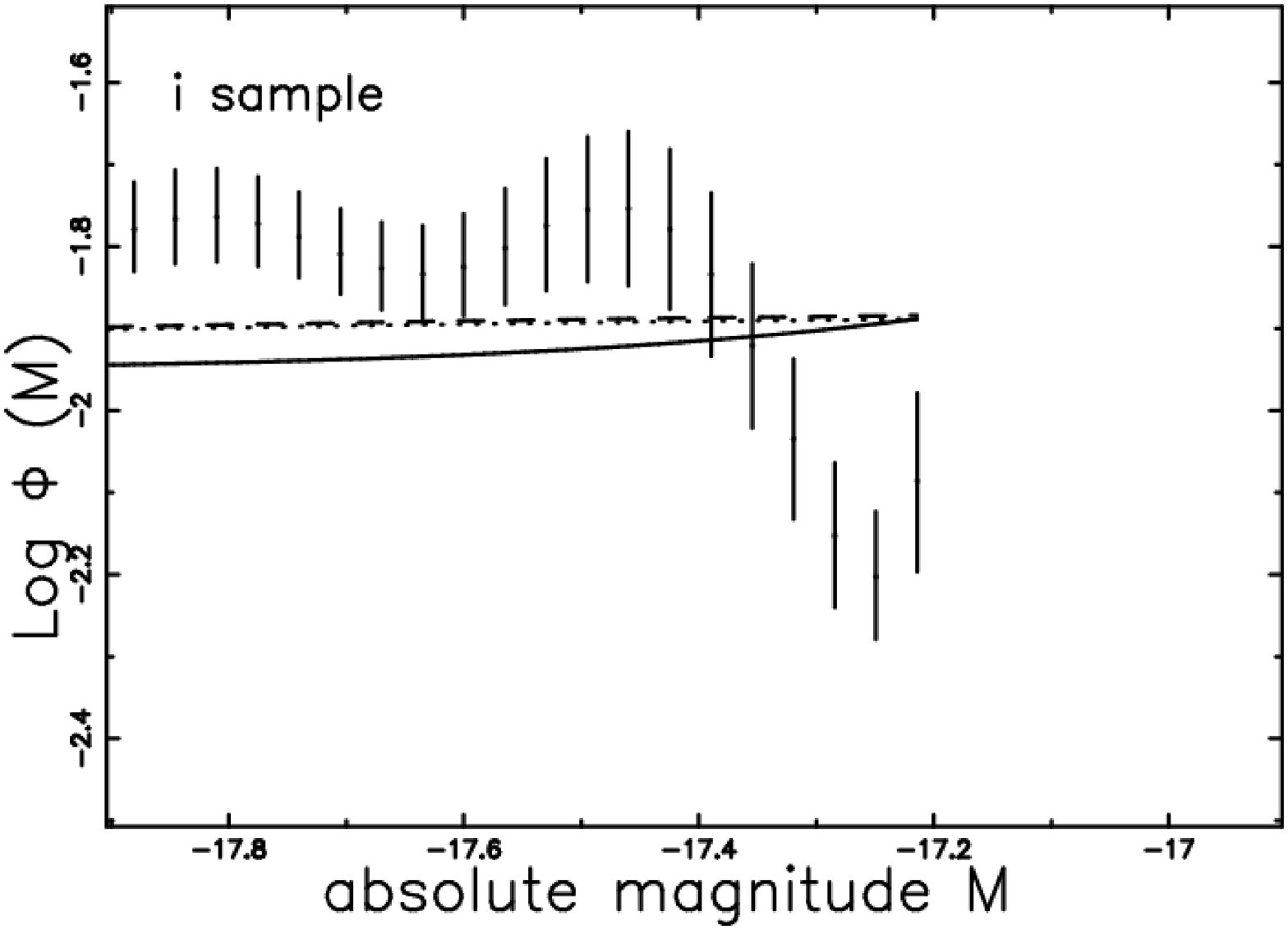}
\caption {The luminosity function data of 
SDSS($i^*$) 
in the low luminosity region 
are represented with error bars. 
The continuous line fit represents our LF~(\ref{pdf5magni}),
the dotted line represents the Schechter function as 
given by our data
and
the dashed
line represents the Schechter function as given 
by \cite{Blanton_2003}.
 }
 \label{due_low_i} 
 \end{figure}

\begin{figure}
 \centering
\includegraphics[width=10cm]{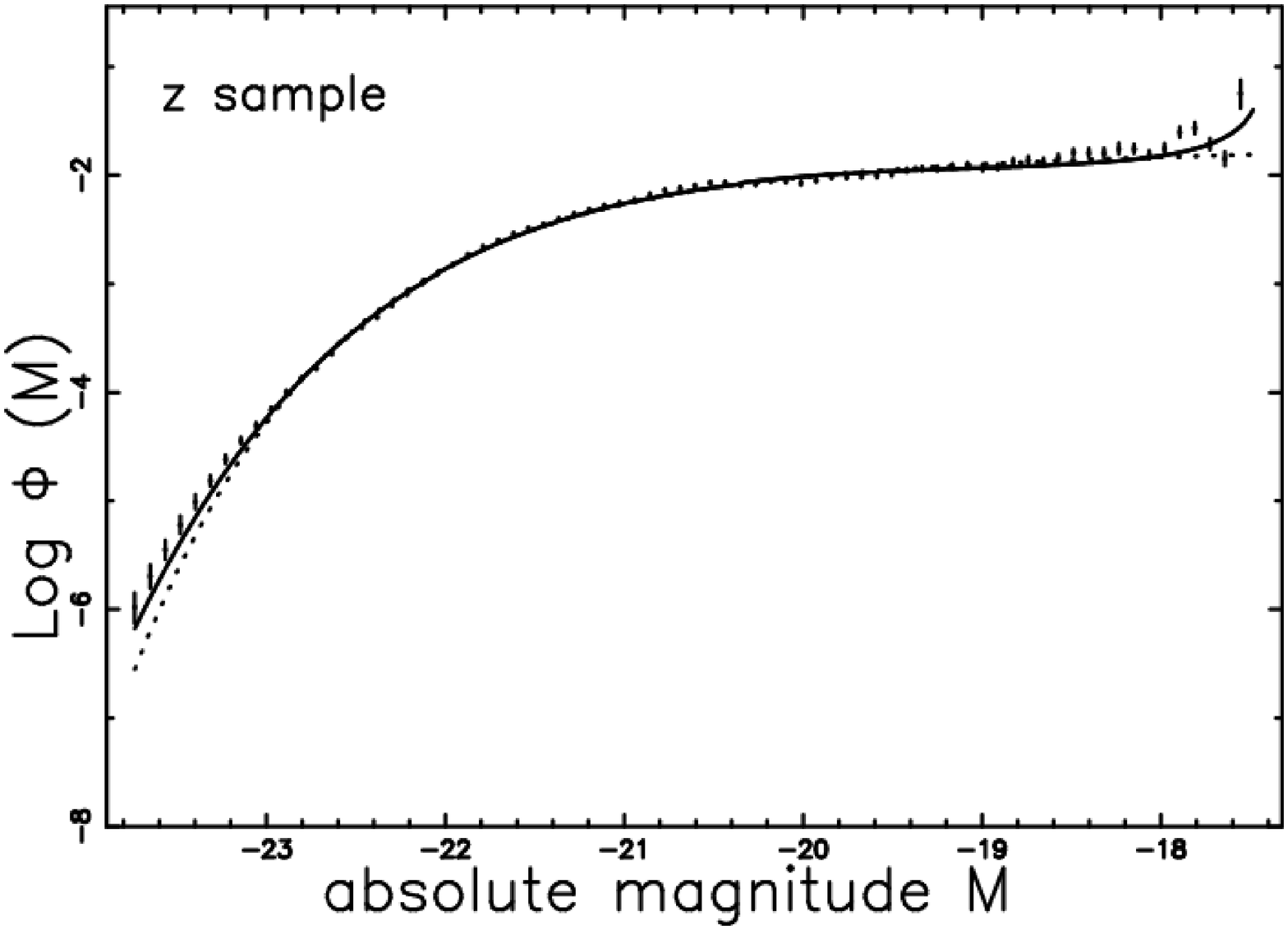}
\caption {The luminosity function data of 
SDSS($z^*$) are represented with error bars. 
The continuous line fit represents our LF~(\ref{pdf5magni}) 
and the dotted 
line represents the Schechter function.
 }
 \label{due_z} 
 \end{figure}

\begin{figure}
 \centering
\includegraphics[width=10cm]{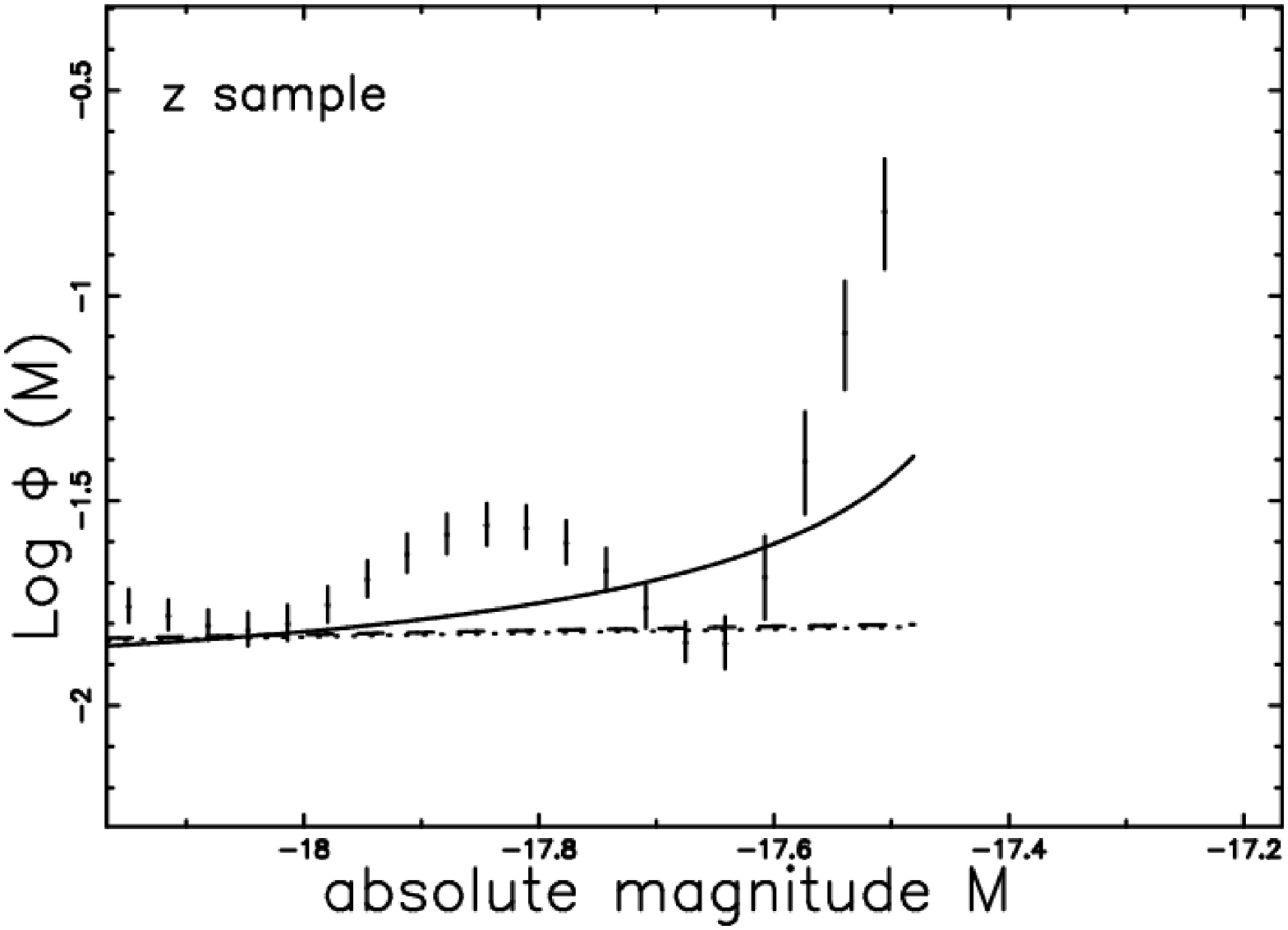}
\caption {The luminosity function data of 
SDSS($z^*$) 
in the low luminosity region 
are represented with error bars. 
The continuous line fit represents our LF~(\ref{pdf5magni}),
the dotted line represents the Schechter function as 
given by our data
and
the dashed
line represents the Schechter function as given 
by \cite{Blanton_2003}.
 }
 \label{due_low_z} 
 \end{figure}

The Schechter function, the new five parameter function 
represented by formula~(\ref{pdf5magni})
and the data are
reported in 
Figure~\ref{due_u},
Figure~\ref{due_low_u},
Figure~\ref{due_g},
Figure~\ref{due_low_g},
Figure~\ref{due_r},
Figure~\ref{due_low_r},
Figure~\ref{due_i},
Figure~\ref{due_low_i},
Figure~\ref{due_z},
and Figure~\ref{due_low_z}
where bands $u^*$, $g^*$, $r^*$, $i^*$ and $z^*$ 
are considered.

The flexiblity of the generalized gamma with five parameters
to fit a sudden increase in the 
low luminosity region is clearly visible in 
Figure~\ref{due_low_z}.

\subsection{The generalized gamma distribution with four parameters}

We can start from equation~(\ref{base5}),
inserting $a=0$, $x=L$ and $b=L^*$: 
\begin{equation}
\Psi(L;L^*,c,k,\Psi^*)=
\Psi^*\frac 
{
{\it k}\, \left( {\frac {L}{L^*}} \right) ^{c{\it k}-1}{{\rm e}^{-
 \left( {\frac {L}{L^*}} \right) ^{{\it k}}}}
}
{
L^*\Gamma \left( c \right) 
}
\quad .
\label{lf4}
\end{equation}

The mathematical range of existence is $ 0 \leq L < \infty $
and the number of parameters is four because $a=0$ and
$\Psi^*$ have been added. 
The averaged luminosity is 
\begin{equation}
{ \langle L \rangle } 
=
\frac
{
{\it L^*}\,\Gamma \left( {\frac {1+ck}{k}} \right) 
}
{
\Gamma \left( c \right) 
}
\quad ,
\label{lmedio4}
\end{equation}
and the mode is
at 
\begin{equation}
L=
\left( {\frac {ck-1}{k}} \right) ^{\frac{1}{k}} {\it L^*}
\quad . 
\label{mode4} 
\end{equation}

The magnitude version of the LF is
\begin{equation}
\Psi (M) dM =
\frac
{
{ \Psi^*}\,0.4\,\ln \left( 10 \right)\,k{10}^{- 0.4\,ck \left( M- \,{ M^*}
 \right) }{{\rm e}^{- \,{10}^{- 0.4\, \left( M- \,{ 
M^*} \right) k}}} 
}
{
\Gamma \left( c \right) 
}
 dM \quad.
\label{pdf4magni}
\end {equation}
The mode when expressed in magnitude is 
at 
\begin{equation}
M = -\frac 
{
 1.0857\,\ln \left( {\frac {c{\it k}- 1}{{\it k}}} \right) 
}
{
k
} 
+ {\it M^*}
\quad .
\label{mode4magni}
\end{equation}

This function contains the four parameters $c$,
k$k$, $M^*$
and $\Psi^*$ and the numerical values obtained 
are reported in Table~\ref{dataLF4}.

 \begin{table}
 \caption[]{
 Parameters for fits to LF
 in SDSS Galaxies with the 
 four parameter function represented by 
 formula (\ref{pdf4magni}).

}
 \label{dataLF4} 
 \[
 \begin{array}{lccccc}
 \hline
parameter    &  u^*   &  g^* & r^*  & i^* & z^*  \\ \noalign{\smallskip}
 \hline
 \noalign{\smallskip}
M^* - 5\log_{10}h   
 & -17.34 
 & -19.43  
 & -20.25 
 & -20.29 
 & -20.76 
 \\
\Psi^* [h^3~Mpc^{-3}] 
 & 0.101 
 & 0.231  
 & 0.65   
 & 0.097  
 & 0.2  
 \\
c & 0.47 
 & 0.08  
 & 0.02   
 & 0.247  
 & 0.10   
 \\ 
k 
 & 0.842  
 & 1.019  
 & 0.93  
 & 0.839  
 & 0.864 
 \\ 
\chi^2 
 & 283 
 & 747 
 & 2188
 & 1866
 & 2915
\\
\chi_{red}^2 
 & 0.591 
 & 1.256
 & 3.266
 & 2.648
 & 3.991
 \\
AIC~k=4 
 & 291
 & 755 
 & 2196
 & 1874
 & 2921
\\
BIC~k=4 
 & 307
 & 773 
 & 2214
 & 1893
 & 2921
\\
\chi^2 Schechter
 & 330 
 & 753 
 & 2260
 & 2282 
 & 3245
\\
\chi_{red}^2 -Schechter 
 & 0.689
 & 1.263
 & 3.368
 & 3.232
 & 4.403
\\

\chi^2 Blanton~2003 
 & 341 
 & 756 
 & 2276
 & 2283 
 & 3262
\\
 \end{array}
 \]
 \end {table}

\section{The generalized gamma distribution for mass }

\label{masses}
We assume that the masses of galaxies, 
 $\mathcal {M}$,
are distributed as a generalized gamma.
We can start from equation~(\ref{base5})
inserting $a=0$ , $x= \mathcal{M} $ and $b=\mathcal{M}^*$ 
\begin{equation}
\Psi( \mathcal{M} ;\mathcal{M}^*,c,k,\Psi^*)=
\Psi^*\frac 
{
{\it k}\, \left( {\frac {\mathcal{M} }{\mathcal{M}^* }} \right) ^{c{\it k}-1}{{\rm e}^{-
 \left( {\frac {\mathcal{M}}{\mathcal{M}^*}} \right) ^{{\it k}}}}
}
{
\mathcal{M}^*\Gamma \left( c \right) 
}
\quad.
\label{ml4}
\end{equation}
This is a generalized gamma distribution with 
a scale parameter ${\mathcal M^*}$,
 {\it c} and {\it k} are shape parameters .
The mathematical range of existence is $ 0 \leq \mathcal {M} < \infty $
and the number of parameters is four because 
$\Psi^*$ has been added.
 The average value
is
\begin{equation}
{\langle{\mathcal M}\rangle}= \Psi^*{\mathcal M^*}
\frac{
\Gamma \left( {\frac {1+ck}{k}} \right) 
}
{
\Gamma \left( c \right) 
}
\quad.
\label{massamedia}
\end{equation}
We now follow the pattern fixed in \cite{Zaninetti2008} 
where the PDF representing the mass of galaxies
was assumed to be a Kiang function, see \cite {kiang}.
The mass-luminosity relationship 
is assumed to follow a power law 
having the form 
\begin{equation}
\frac {\mathcal M }{\mathcal M^*} = \left ( \frac {L }{L^*}
\right ) ^{\frac{1}{d} }\quad,
\label{massa}
\end{equation}
where $1/d$ is an exponent which connects the mass to
luminosity.
The PDF (\ref{ml4} ) is therefore transformed into
 the
following form:
\begin{equation}
\Psi (L) dL =\Psi^*
\frac
{
k \left( {\frac {L}{{\it L^*}}} \right) ^{\frac{ck-1}{d}} 
{\rm e}^{- \left( {\frac {L}{ L^*}} \right) ^{\frac{k}{d}}}
\left( {\frac {L}{{\it L^*}}} \right) ^{\frac{1}{d}} 
}
{
d \,\Gamma \left( c \right) L
}
 , \label{ml5}
\end {equation}

The mathematical range of existence is $ 0 \leq L < \infty $
and the averaged luminosity, $ { \langle L \rangle } $, is 
\begin{equation}
{ \langle L \rangle } 
=
\Psi^*
\frac
{
{\it L^*}\,\Gamma \left( {\frac {d+ck}{k}} \right) 
}
{
\Gamma \left( c \right) 
}
\quad.
\end{equation}
The magnitude version of the previous LF is: 
\begin{equation}
\Psi (M) dM =
{\it \Psi^*} 
\frac
{
 0.4\, {\it ak}\,{10}^{- 0.4\,{\frac {c
{\it ak}\, \left( - \,{\it M^*}+{\it M} \right) }{d}}}{{\rm e}^
{- {10}^{- 0.4\,{\frac { \left( - \,{\it M^*}+{
\it M} \right) {\it ak}}{d}}}}}\ln \left( 10 \right) 
}
{
\Gamma \left( c \right) d
}
\quad.
\label{pdfml5} 
\end{equation} 
The mode expressed in magnitude is at 
\begin{equation}
M = 
M^* - 1.0857\,{\frac {\ln \left( c \right) d}{{
\it k}}}
\quad.
\label{modeml5magni}
\end{equation}

This function contains the five parameters 
$c$,
$k$, 
$d$, 
$M^*$
and $\Psi^*$ and the numerical values obtained 
are reported in Table~\ref{dataMLLF5}.

 \begin{table}
 \caption[]
 {
 Parameters of fits to LF
 in SDSS Galaxies of the 
 five parameter function represented by 
 formula (\ref{pdfml5}) 
 deduced from the mass-luminosity 
 relationship.
 \label{dataMLLF5} 
 }
 \[
 \begin{array}{lccccc}
 \hline
parameter    &  u^*   &  g^* & r^*  & i^* & z^*  \\ \noalign{\smallskip}
 \hline
 \noalign{\smallskip}

M^* - 5\log_{10}h  
 & -17.34  
 & -19.39  
 & -20.18 
 & -20.43 
 & -20.53 
 \\
\Psi^* [h^3~Mpc^{-3}] 
 & 0.101 
 & 0.19  
 & 0.33  
 & 0.12  
 & 0.11 
 \\
c & 0.47 
 & 0.1  
 & 0.05  
 & 0.17 
 & 0.22  
 \\ 
k & 0.855  
 & 1.029  
 & 0.93 
 & 0.88  
 & 0.824  
 \\ 
d & 1.015  
 & 1.024  
 & 1.023  
 & 1.018  
 & 1.017 
 \\ 
\chi^2 
 & 283 
 & 751 
 & 2217
 & 1896
 & 3009
\\
\chi_{red}^2 
 & 0.592 
 & 1.265
 & 3.313
 & 2.693
 & 3.313
 \\
AIC~k=3 
 & 289
 & 757 
 & 2223
 & 1902
 & 3015
\\
 BIC~k=3 
 & 314
 & 783 
 & 2249
 & 1929
 & 3042
\\

\chi^2 ~Schechter 
 & 330
 & 753 
 & 2260
 & 2282 
 & 3245
\\
\chi_{red}^2 -Schechter 
 & 0.689
 & 1.263
 & 3.368
 & 3.232
 & 4.403
\\

\chi^2 ~Blanton~2003 
 & 341 
 & 756 
 & 2276
 & 2283 
 & 3262
\\
 \end{array}
 \]
 \end {table}

From a careful analysis of Table~\ref{dataMLLF5}, 
it is possible to conclude that $1.015 < d < 1.024$.
Because we have assumed that $L\propto \mathcal {M} ^d$,
the nonlinear relationship between mass and luminosity 
is a weak effect.
A weak dependence of the parameter 
that regulates mass and luminosity
has also been found in \cite{Zaninetti2008b} when the PDF
which represents the mass of galaxies is 
the product of two gamma variates with argument 2.

\section{The Schechter function with transformation of location}

\label{location}
The Schechter LF of galaxies 
once the location $L_a$ is introduced 
is: 
\begin{equation}
\Phi (L) dL = (\frac {\Phi^*}{L^*}) (\frac {(L-L_a)}{L^*})^{\alpha}
\exp \bigl ( {- \frac {(L-L_a)}{L^*}} \bigr ) dL \quad ,
\label{schechter_loc4} 
\end {equation}
and is characterized by four parameters.

The average luminosity, $ { \langle L \rangle } $, is 
\begin{equation}
{ \langle L \rangle } 
=
\Phi^*( L^*\,\Gamma \left( 2+\alpha \right) +L_a\,
\Gamma \left( \alpha+1 \right)) 
\quad ,
\end{equation}
and the mode is
at 
\begin{equation}
L=
\alpha\,L^*+L_a
\quad , 
\end{equation}
or
\begin{equation}
M =
- 1.085\,\ln \left( \alpha\,{{\rm e}^{- 0.921\,{\it 
M^*}}}+{{\rm e}^{- 0.921\,{\it M_a}}} \right)
\quad .
\label{modem} 
\end{equation}
The magnitude version of this LF is
\begin{eqnarray}
\Phi (M) dM =
0.4\,\Phi^*\, \left( {10}^{ 0.4\,M^*- 0.4\,M
}- \,{10}^{ 0.4\,M^*- 0.4\,M_a} \right) ^{
\alpha}
\nonumber \\
{{\rm e}^{- \,{10}^{ 0.4\,M^*- 0.4\,
M}+{10}^{ 0.4\,M^*- 0.4\,M_a}}}{10}^{
 0.4\,M^*- 0.4\,M}\ln \left( 10 \right) 
\quad. 
\label{equationm_schechter_location}
\end{eqnarray}
 
This function contains the four parameters $\alpha$,
$M_a$, $M^*$
and $\Phi^*$ and the numerical values obtained 
are reported in Table~\ref{schlocation}.
 \begin{table}
 \caption[]
 {
 Parameters of fits to LF
 in SDSS Galaxies of the 
 four parameter 
 Schechter function with location
 as represented by 
 formula (\ref{equationm_schechter_location}).
 }
 \label{schlocation} 
 \[
 \begin{array}{lccccc}
 \hline
parameter    &  u^*   &  g^* & r^*  & i^* & z^*  \\ \noalign{\smallskip}
 \hline
 \noalign{\smallskip}
 M^* - 5\log_{10}h  
 & -17.92 
 & -19.37  
 & -20.41 
 & -20.83 
 & -21.15 
 \\
\Phi^* [h^3~Mpc^{-3}] 
 & 0.03    
 & 0.021  
 & 0.015  
 & 0.014  
 & 0.013 
 \\
\alpha 
 & -0.9 
 & -0.84 
 & -0.99 
 & -1.02 
 & -1.01 
 \\ 
M_a ~ [mags] 
  & -12.1 
  & -14.87 
 & -15.14  
 & -12.72   
 & -16.39   
 \\ 
\chi^2 
 & 333 
 & 723 
 & 2108
 & 2453
 & 3133
\\
\chi_{red}^2 
 & 0.69 
 & 1.21
 & 3.14
 & 3.47
 & 4.25
 \\

AIC~k=4 
 & 341
 & 731 
 & 2116
 & 2451
 & 3141
\\
 BIC~k=4 
 & 358
 & 748 
 & 2134
 & 2479
 & 3159
\\
\chi^2~ Schechter 
 & 330 
 & 753 
 & 2260
 & 2282 
 & 3245
\\
\chi_{red}^2 -Schechter 
 & 0.689
 & 1.263
 & 3.368
 & 3.232
 & 4.403
\\
\chi^2~Blanton~2003 
 & 341 
 & 756 
 & 2276
 & 2283 
 & 3262
\\
 \end{array}
 \]
 \end {table}

From a careful analysis of Table~\ref{schlocation}, 
it is possible to conclude that in three cases out of five analyzed,
the Schechter LF with transformation of location produces 
a "better fit" than the standard version.

\section{The radial distribution of galaxies}
\label{secz}
Some useful formulae connected with the 
Schechter LF, see formula~(\ref{equation_schechter}),
in a Euclidean, non-relativistic 
and homogeneous universe 
are reviewed;
by analogy, new formulae for 
the generalized gamma distribution with four parameters,
see formula~(\ref{lf4}), are derived.

\subsection{Dependence of the LF on $z$}

The radiation flux, {\it f}, is introduced 
\begin{equation}
f = \frac{L}{4 \pi r^2} 
\quad,
\end{equation}
where {\it r} represents the distance of the galaxy.
The joint distribution in redshift, $z$,
and $f$ adopting the Schechter LF,
see formula~(1.104) in~\cite{pad} or formula~(1.117) 
in~\cite{Padmanabhan_III_2002},
 is:
\begin{equation}
\frac{dN}{d\Omega dz df} = 
4 \pi \bigl ( \frac {c_L}{H_0} \bigr )^5 z^4 \Phi (\frac{z^2}{z_{crit}^2})
\label{nfunctionz} 
\quad,
\end {equation}
where $d\Omega$, $dz$ and $ df $ 
represent the differential of
the solid angle, 
the redshift and the flux, respectively.
The critical value of $z$, $z_{crit}$,
 is 
\begin{equation}
 z_{crit}^2 = \frac {H_0^2 L^* } {4 \pi f c_L^2}
\quad,
\end{equation} 
where $c_L$ is the velocity of light and
$H_0$ is the Hubble constant.

The number of galaxies, $N_S(z,f_{min},f_{max})$ 
comprised between a minimum value of flux,
 $f_{min}$, and maximum value of flux $f_{max}$,
for the Schechter LF
can be computed through the following integral: 
\begin{equation}
N_S (z) = \int_{f_{min}} ^{f_{max}}
4 \pi \bigl ( \frac {c_L}{H_0} \bigr )^5 z^4 \Phi (\frac{z^2}{z_{crit}^2})
df
\quad.
\label{integrale_schechter} 
\end {equation}
This integral does not have an analytical solution 
and we must perform 
a numerical integration.

The number of galaxies 
for the Schechter LF
in $ z$ and $ f$ as given by 
formula~(\ref{nfunctionz}) has a maximum 
at $z=z_{pos-max}$,
where 
\begin{equation}
 z_{pos-max} = z_{crit} \sqrt {\alpha +2 }
\quad,
\end{equation} 
which can be re-expressed as
\begin{equation}
 z_{pos-max} =
\frac
{
\sqrt {2+\alpha}\sqrt {{10}^{ 0.4\,{\it M_{\sun}}- 0.4\,{\it M^*}}}{
\it H_0}
}
{
2\,\sqrt {\pi }\sqrt {f}{\it c_L}
}
\quad,
\label{zmax_sch}
\end{equation}
where $M_{\sun}$ is the reference magnitude 
of the sun in the bandpass under consideration.
For the sake of clarity, 
we report 
the generalized gamma LF of galaxies 
with four parameters (equation~(\ref{lf4})),
in the following LF4
\begin{equation}
\Psi(L;L^*,c,k,\Psi^*)=
\Psi^*\frac 
{
{\it k}\, \left( {\frac {L}{L^*}} \right) ^{c{\it k}-1}{{\rm e}^{-
 \left( {\frac {L}{L^*}} \right) ^{{\it k}}}}
}
{
L^*\Gamma \left( c \right) 
}
\quad.
\label{lf44}
\end{equation}
The joint distribution in $z$ and $f$ 
of the generalized gamma LF4 of galaxies 
is 
\begin{equation}
\frac{dN}{d\Omega dz df} = 
4 \pi \bigl ( \frac {c_L}{H_0} \bigr )^5 z^4 \Psi (\frac{z^2}{z_{crit}^2})
\label{nfunctionz_mia} 
\quad.
\end {equation}
The number of galaxies, $N_{LF4}(z,f_{min},f_{max})$ 
of the LF4 comprised between 
 $f_{min}$ and $f_{max}$,
can be computed through 
the following integral: 
\begin{equation}
N_{LF4} (z) = \int_{f_{min}} ^{f_{max}}
4 \pi \bigl ( \frac {c_L}{H_0} \bigr )^5 z^4 \Psi (\frac{z^2}{z_{crit}^2})
df 
\quad,
\label{integrale_mia} 
\end {equation}
and in this case again 
a numerical integration must be performed.

The number of galaxies
of the LF4
 as given 
by formula~(\ref{nfunctionz_mia}) 
has a maximum at 
$z_{pos-max}$ where 
\begin{equation}
 z_{pos-max} = 
{e^{1/2\,{\frac {\ln \left( 1+c{\it k} \right) -\ln \left( {\it k}
 \right) }{{\it k}}}}}{\it z_{crit}}
\quad,
\end{equation} 
which can be re-expressed as
\begin{equation}
 z_{pos-max} = 
\frac
{
{e^{1/2\,{\frac {\ln \left( 1+{\it c}\,{\it k} \right) -\ln 
 \left( {\it k} \right) }{{\it k}}}}}\sqrt {{10}^{ 0.4\,{\it 
{\it M_{\sun}}}
- 0.4\,{\it M^*}}}{\it H_0k}
}
{
2\,\sqrt {\pi }\sqrt {f}{\it c_L}
}
\quad.
\label{zmax_mia}
\end{equation} 
The formulae previously derived are now tested 
on the 
2dFGRS catalogue
available at the web site: http://msowww.anu.edu.au/2dFGRS/.
The 2dFGRS catalogue contains redshifts for
221,414 galaxies
 brighter
than a magnitude limit of bJmag=19.45.
The galaxies cover
an area of approximately 1500 square degrees  
and  more details can be found in  \cite{Colless2001}.
In particular we added together the file parent.ngp.txt which 
contains 145,652 entries for NGP strip sources and 
the file parent.sgp.txt which 
contains 204,490 entries for SGP strip sources.
Once the heliocentric redshift had been selected, 
we processed 219,107 galaxies with 
$0.001 \leq z \leq 0.3$.
The parameters of the Schechter function for 2dFGRS 
are reported in Table~\ref{parameters},
see first line of Table~3 in \cite{Madgwick_2002}.

It is interesting to point out that other values 
for $h$ 
different from 1
shift all the absolute magnitudes by 
$5\log_{10}h$ and change the
number densities by a factor $h^3$.

\begin{table}
 \caption[]{Parameters of the Schechter function for 
 2dFGRS. }
 \label{parameters}
 \[
 \begin{array}{lc}
 \hline
 \hline
 \noalign{\smallskip}
parameter & 2dFGRS \\ \noalign{\smallskip}
M^* - 5\log_{10}h ~ [mags] & ( -19.79 \pm 0.04)+5 Log h \\ \noalign{\smallskip}
\alpha & -1.19 \pm 0.01 \\ \noalign{\smallskip}
\Phi^* ~h^3~[Mpc^{-3}] & ((1.59 \pm 0.1)10^{-2})\times h^3 \\ \noalign{\smallskip}
 \hline
 \hline
 \end{array}
 \]
 \end {table}

\begin{table} 
 \caption[]{Parameters of
 LF4,
  formula~(\ref{nfunctionz_mia}), \\
 based on the 2dFGRS data 
 (triplets generated by the author) } 
 \label{para_physical} 
 \[ 
 \begin{array}{lc} 
 \hline 
~ & 2dFGRS \\ \noalign{\smallskip} 
 \hline 
 \noalign{\smallskip} 
c & 0.016 \pm0.01 \\ \noalign{\smallskip}
M^* - 5\log_{10}h [mags] & -19.15 \pm 0.029 \\ \noalign{\smallskip}
\Psi^* [h^3~Mpc^{-3}] & 2.24 \pm 1.42 \\ \noalign{\smallskip}
k & 0.82 \pm 0.06 \\ \noalign{\smallskip} 
 \hline 
 \hline 
 \end{array} 
 \] 
 \end {table}

\begin{figure}
\begin{center}
\includegraphics[width=10cm]{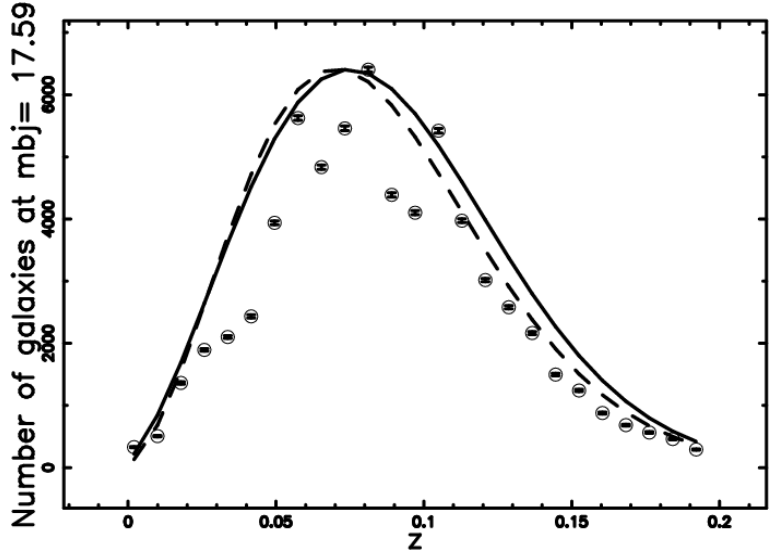}
\end {center}
\caption{
The galaxies of the 2dFGRS database with 
$ 16.77 \leq bJmag \leq 18.40 $ or 
$ 4677 \frac {L_{\sun}}{Mpc^2} \leq 
f \leq 21087 \frac {L_{\sun}}{Mpc^2}$
(with $bJmag$ representing the 
relative magnitude used in object selection),
are isolated 
in order to represent a chosen value of $m$ 
and then organized as frequency versus
heliocentric redshift,
 (empty circles);
the error bar is given by the square root of the frequency.
The maximum in the frequencies of observed galaxies is 
at $z=0.085$ when $\mathcal{M_{\sun}}$ = 5.33 and
$h$=1 .
The theoretical curve generated by
the Schechter function of luminosity 
(formula~(\ref{nfunctionz}) and parameters
as in column 2dFGRS of Table~\ref{parameters}) 
is drawn (full line).
The theoretical curve generated by
LF4,
 formula~(\ref{nfunctionz_mia}),
and parameters as in column 
2dFGRS of Table~\ref{para_physical})
is drawn (dashed line);
 $\chi^2$= 8078 for the Schechter function and $\chi^2$= 6654
for LF4.
}
 \label{maximum_flux}%
 \end{figure}

\begin{figure}
\begin{center}
\includegraphics[width=10cm]{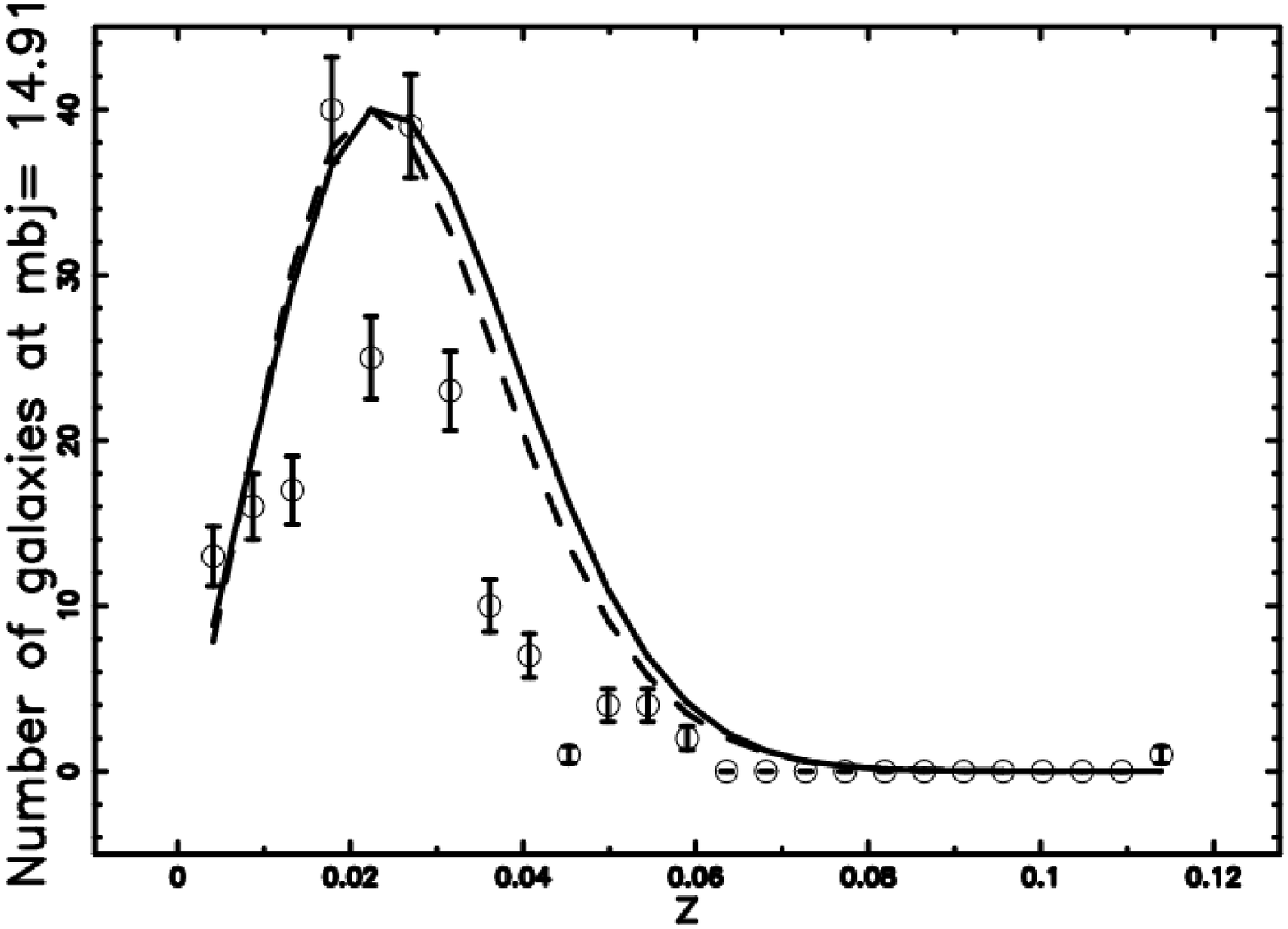}
\end {center}
\caption{
Galaxies in the 2dFGRS with 
$ 14.79 \leq bJmag \leq 15.02 $
or 
$ 105142 \frac {L_{\sun}}{Mpc^2} \leq 
f \leq 129757 \frac {L_{\sun}}{Mpc^2}$.
The maximum in the frequencies of observed galaxies 
is at $z=0.02$,
$\mathcal{M_{\sun}}$ = 5.33, 
$\chi^2$= 348 for the Schechter function (full line)
and $\chi^2$= 243
for
LF4
(dashed line).
}
 \label{maximum_flux_2}%
 \end{figure}

Figure~\ref{maximum_flux}
and Figure~\ref{maximum_flux_2}
report the number of observed galaxies
in the 2dFGRS catalogue for 
two different apparent magnitudes and 
the theoretical curves as represented by 
formula~(\ref{nfunctionz}) 
and formula~(\ref{nfunctionz_mia}).

\begin{figure}
\begin{center}
\includegraphics[width=10cm]{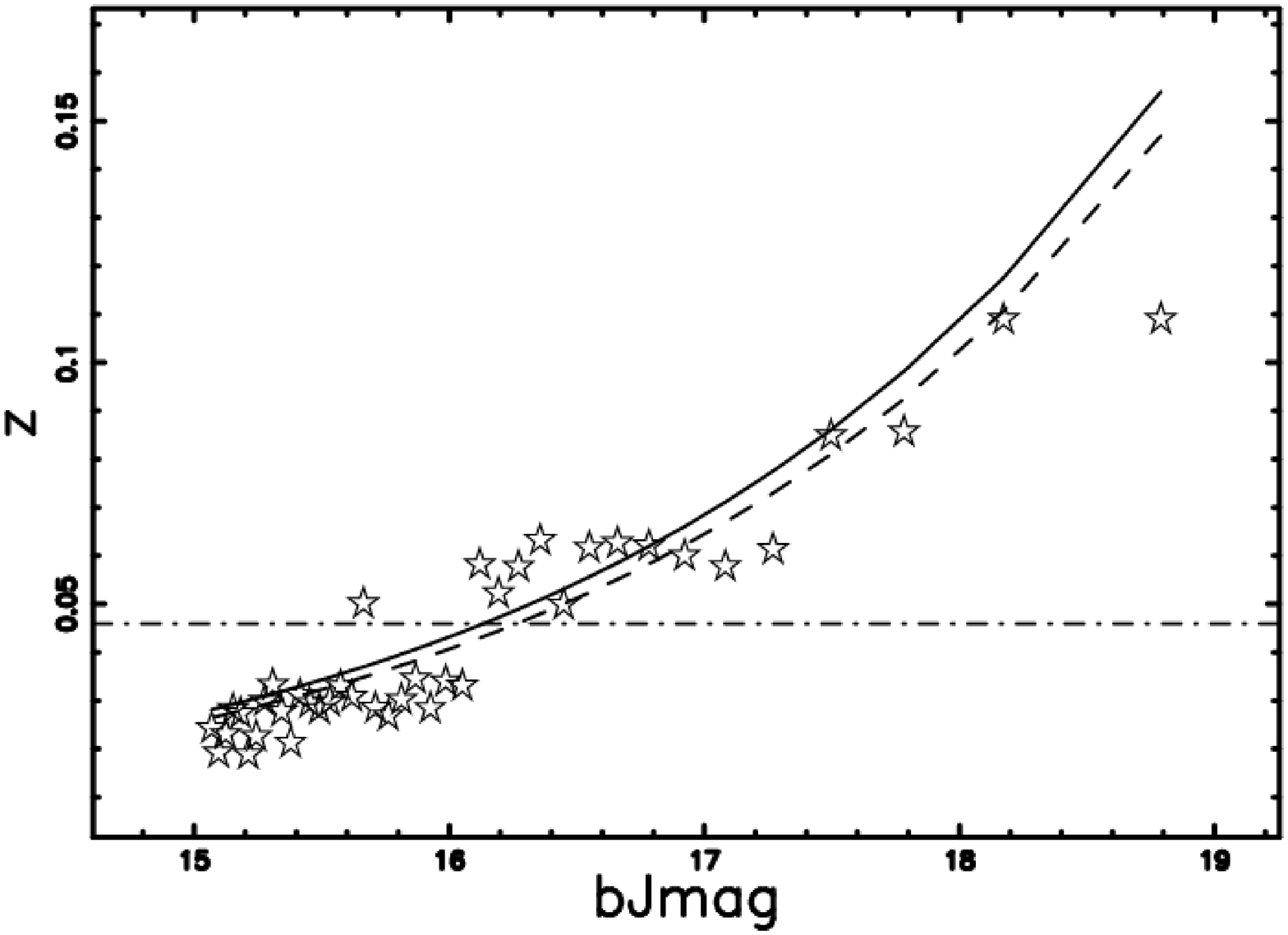}
\end {center}
\caption{
Value of $z$ at which the number of
 galaxies in the 2dFGRS database
is maximum as a function of 
the apparent magnitude $ bJmag$ 
(stars),
theoretical curve of the maximum for the 
Schechter function as represented by formula~(\ref{zmax_sch}) 
(full line) and
theoretical curve of the maximum for 
LF4
as represented by 
formula~(\ref{zmax_mia}) 
(dashed line) 
when $\mathcal{M_{\sun}}$ = 5.33 and
$h$=1. 
The dash-dot-dash horizontal line represents the upper limit
of the complete sample.
}
 \label{zeta_max_flux}%
 \end{figure}

Due to the importance of the maximum as a function of $z$ in the number of 
galaxies, Figure~\ref{zeta_max_flux} reports 
the observed histograms in the 2dFGRS database
and the theoretical curves as a function of magnitude.

\begin{figure}
\begin{center}
\includegraphics[width=10cm]{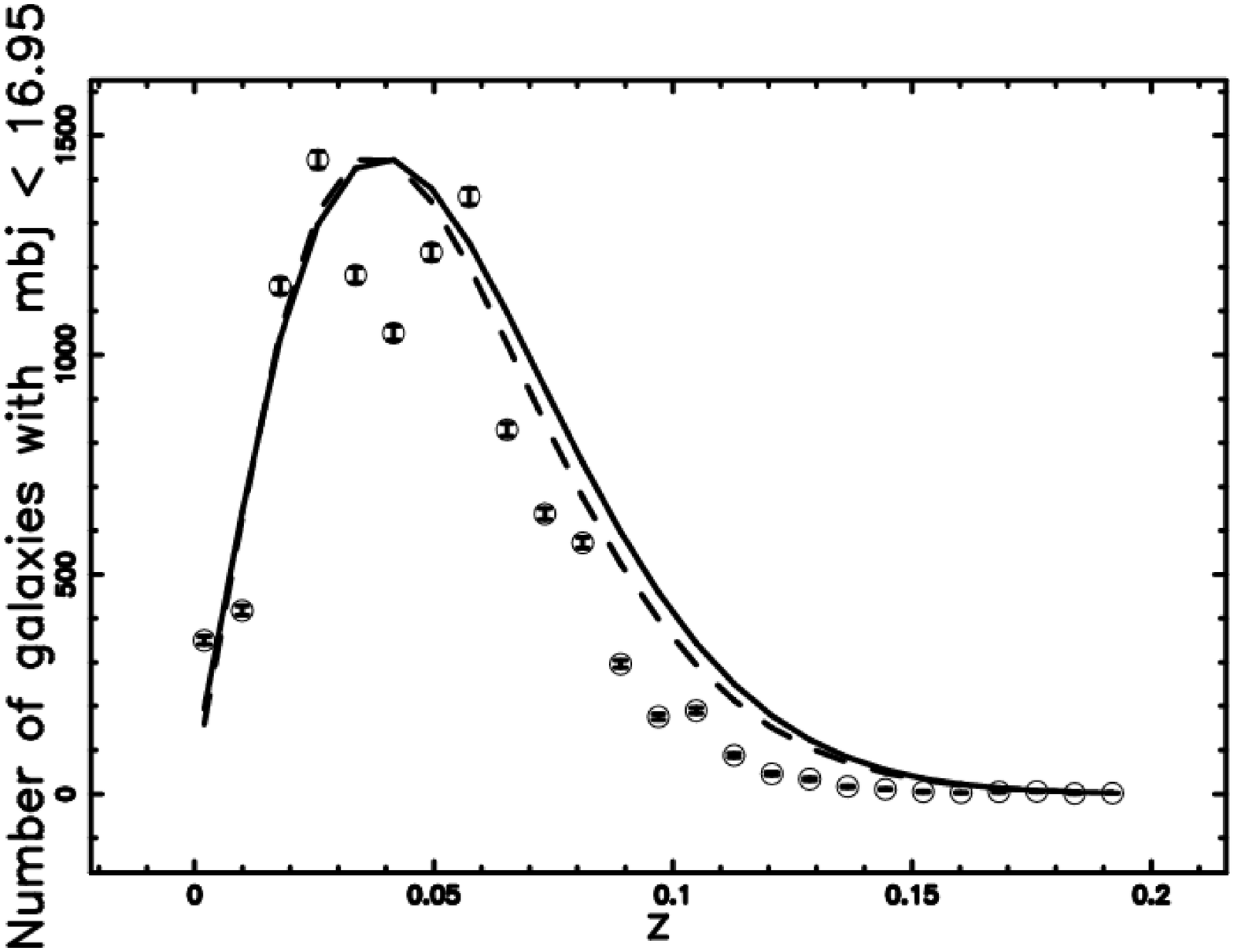}
\end {center}
\caption{The galaxies in the 2dFGRS database with 
$ 13.44 \leq bJmag \leq 16.94 $
or 
$ 17950 \frac {L_{\sun}}{Mpc^2} \leq 
f \leq 493844 \frac {L_{\sun}}{Mpc^2}$,
are organized as frequency versus
heliocentric redshift, (empty stars).
The theoretical curves generated by
the integral of the Schechter function in flux 
(formula~(\ref{integrale_schechter}) with parameters
as in Table~\ref{parameters}) 
(full line)
and by the integral of 
LF4
as represented by 
formula~(\ref{integrale_mia}) 
with parameters as in column 2dFGRS of Table~\ref{para_physical})
(dashed line) are drawn
when $\mathcal{M_{\sun}}$ = 5.33 and
$h$=1.
The numerical analysis gives 
$\chi^2$= 3314 for the Schechter function (full line)
and $\chi^2$= 2246 
for
LF4
(dashed line).
}
 \label{maximum_flux_all}%
 \end{figure}

The total number of galaxies in the 2dFGRS database
is reported in Figure~\ref{maximum_flux_all} as well as
the theoretical curves as represented 
by the numerical integration of formula~(\ref{nfunctionz})
and formula~(\ref{nfunctionz_mia}).

\begin{figure}
 \centering
\includegraphics[width=10cm]{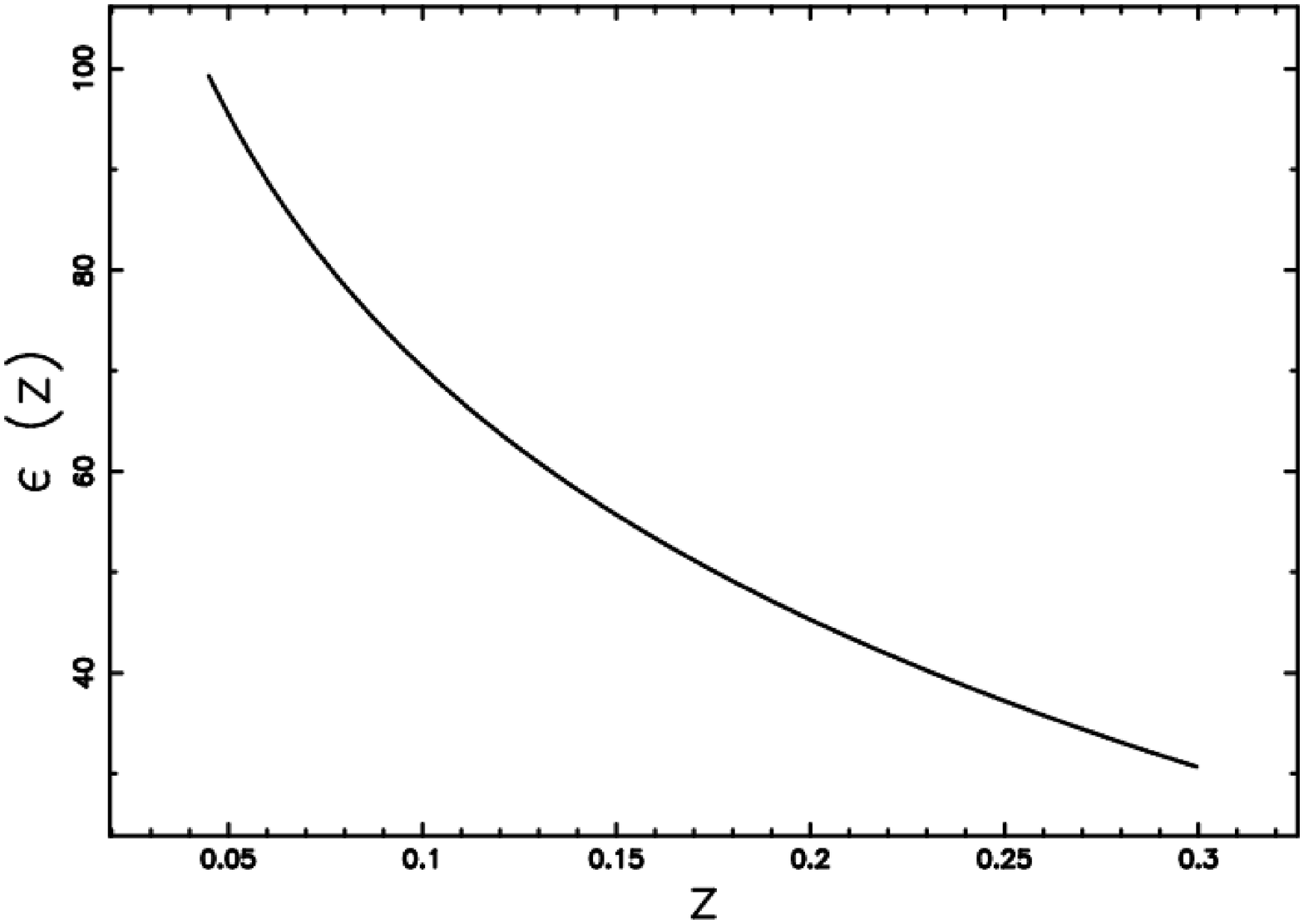}
\caption {
Range in percent of the observable absolute 
magnitude as a function of the redshift
when 
$M_{max}$=-16, 
$M_{min}$= -22,
$m_L$=19.61 
and
$h$=1.
 }
 \label{completo} 
 \end{figure}

Particular attention should be paid to the concept
of limiting magnitude and to the corresponding 
completeness in absolute magnitude of 
the considered catalogue as a function of redshift.
The observable absolute magnitude 
as a function of the limiting apparent magnitude,
 $m_L$,
is 
\begin{equation}
M_L = 
m_{{L}}-5\,{\it \log_{10}} \left( {\frac {{\it c_L}\,z}{H_{{0}}}}
 \right) -25
\quad .
\end{equation}
The interval covered by the LF of galaxies, 
 $\Delta M $,
is defined as 
\begin{equation}
\Delta M = M_{max} - M_{min}
\quad,
\end{equation}
where $M_{max}$ and $M_{min}$ are the 
maximum and minimum
absolute 
magnitudes of the LF for the considered catalogue.
The real observable interval in absolute magnitude,
 $\Delta M_L $,
 is 
\begin{equation}
\Delta M_L = M_{L} - M_{min}
\quad.
\end{equation}
We can therefore introduce the range
of observable absolute maximum magnitude 
expressed in percent, 
 $ \epsilon(z) $,
as
\begin{equation}
\epsilon(z) = \frac { \Delta M_L } {\Delta M } \times 100
\quad.
\end{equation}
This is a number that represents the 
completeness of the sample and
 can be considered to be a version in percentage terms
of the Malmquist bias for the apparent magnitude 
of a limited number of samples,
see \cite{Malmquist_1920,Malmquist_1922}.
Figure~\ref{completo} shows the behavior of 
the range in absolute magnitude in percent
as a function of $z$ when
the limiting magnitude of
2dFGRS is $m_L$=19.61~.
From the previous figure, it is possible to conclude that 
the 2dFGRS catalogue
is complete for $z\leq0.0442$~.

In this section we have adopted the absolute magnitude of
the sun in the $b_j$ filter 
$\mathcal{M_{\sun}}$ = 5.33, see
\cite{Einasto_2009,Eke_2004}.

\section{Summary}
We have presented three 
new LFs of galaxies 
based on two different versions of the
generalized gamma distribution.
Before continuing, we should spend 
some time discussing the high values 
of $\chi^2$ 
obtained in
Tables~\ref{dataLF5}, \ref{dataLF4}, \ref{dataMLLF5}
and \ref{schlocation}.
We quote the  following phrase
from web site
http://cosmo.nyu.edu/blanton/lf.html:
"Thus, when comparing to theory it would be totally 
inappropriate to rely on
$\chi^2$  from these errors. Such a $\chi^2$ will always 
be very high even for
very successful theories."
From the previous statement we can deduce that 
the high values of $\chi^2$ are part of 
expected results.
The reduced values of $\chi_{red}^2$ conversely give 
acceptable results.

From our numerical analysis we draw the following conclusions:
\begin{itemize}
\item 
 The generalized gamma galaxy LF 
 with five parameters,
 formula (\ref{pdf5magni}),
 produces a "better fit" than
 the Schechter LF in all five bands considered,
 see Table~\ref{dataLF5}.
  As an example, in the $u^*$ band, 
  $\chi^2$ and  $\chi_{red}^2$  are 282 and 0.59 
  against 330 and 0.68 for the Schechter LF.

\item 
 The generalized gamma galaxy LF  
 with four parameters,
 formula (\ref{pdf4magni}),
 produces a "better fit" than
 the Schechter LF in all five bands considered,
 see Table~\ref{dataLF4}.
  As an example, in the $g^*$ band, 
  $\chi^2$ and  $\chi_{red}^2$  are 747 and 1.263 
  against  753 and 1.263 for the Schechter LF.
 Because the number of parameters is 
 less, 
 $\chi^2$, $AIC$ and $BIC$ are higher than in 
 the case of the generalized gamma galaxy LF  
 with five parameters, 
 see Table~\ref{dataLF5}.
\item 
 The LF of galaxies obtained 
 from the generalized gamma PDF for the mass of 
 galaxies assuming $L \propto M^d$,
 formula~(\ref{pdfml5}),
 produces values of $d$ that are slightly greater than one 
 and a  "better fit" than 
 the Schechter LF in all five bands considered,
 see Table~\ref{dataMLLF5}.
  As an example, in the $r^*$ band, 
  $\chi^2$ and  $\chi_{red}^2$  are 2217 and 3.31 
  against  2260 and 3.36 for the Schechter LF.

\end{itemize}

The Schechter LF of galaxies can also be improved by 
incorporating a transformation of location, 
formula~(\ref{schechter_loc4}),
and the numerical analysis 
produces a "better fit" than the standard
 Schechter LF
in three out of the five bands analyzed, 
see Table~\ref{schlocation}.
  As an example, in the $z^*$ band, 
  $\chi^2$ and  $\chi_{red}^2$  are 3133 and 4.25
  against  3245 and 4.4 for the Schechter LF.

One of the four new LFs, more precisely the
generalized gamma with four parameters, 
(equation~(\ref{lf4})),
was tested on the 2dFGRS database and the analysis 
of $\chi^2$ on the histogram of observed frequencies 
versus redshift produces 
$\chi^2$=6654 
for the new LF against     
$\chi^2$=8078 
for the Schechter function.



\end{document}